\documentclass[5p,times,sort&compress]{elsarticle}

\usepackage{amssymb}
\usepackage{amsmath}
\usepackage{graphicx}
\usepackage{hyperref}
\usepackage{booktabs}
\usepackage{float}
\usepackage{longtable}
\usepackage{tabularx}
\usepackage{multirow}
\usepackage{array}
\usepackage{import}
\usepackage{siunitx}
\usepackage[capitalise]{cleveref}
\usepackage{microtype}
\usepackage{amsmath}
\usepackage{mathtools}

\journal{Reliability Engineering \& System Safety}

\usepackage{tabularx,array}
\newcolumntype{C}[1]{>{\centering\arraybackslash}p{#1}}
\newcolumntype{Y}{>{\raggedright\arraybackslash}X}

\begin{document}

\begin{frontmatter}


\title{A Structured Framework for Prioritizing Unsafe Control Actions in STPA: Case Study on eVTOL Operations}

\author[inst1]{Halima El Badaoui\corref{cor1}}
\ead{halima.el-badaoui@warwick.ac.uk}
\cortext[cor1]{Corresponding Author}
\corref{Shufeng Chen}  
\author[inst1]{Shufeng Chen}
\corref{Mariat James Elizebeth}  
\author[inst1]{Mariat James Elizebeth}
\author[inst2]{Takuya Nakashima}
\author[inst1]{Siddartha Khastgir}

\author[inst1]{Paul Jennings}
\affiliation[inst1]{organization={WMG, University of Warwick},
            addressline={6 Lord Bhattacharyya Way}, 
            city={Coventry},
            postcode={CV4 7AL}, 
            country={United Kingdom}}
\affiliation[inst2]{organization={ Graduate School of Frontier Sciences, The University of Tokyo},
            addressline={ 5-1-5 Kashiwanoha }, 
            city={Kashiwa},
            postcode={277-8563}, 
            country={Japan}}


%





\begin{abstract}

Systems Theoretic Process Analysis (STPA) is a widely recommended method for analysing complex system safety. STPA can identify numerous Unsafe Control Actions (UCAs) and requirements depending on the level of granularity of the analysis and the complexity of the system being analysed. Managing numerous results is challenging, especially during a fast-paced development lifecycle. Extensive research has been done to optimize the efficiency of managing and prioritising the STPA results. However, maintaining the objectivity of prioritisation and communicating the prioritised results have become common challenges. In this paper, the authors present a complementary approach that incorporates inputs from both the safety analysts and domain experts to more objectively prioritise UCAs. This is done by evaluating the severity of each UCA, the impact factor of each controller or decision maker that issues the UCA, and the ranking provided by the subject matter experts who assess the UCA criticalities based on different factors. In addition, a Monte Carlo simulation is introduced to reduce subjectivity and relativity, thus enabling more objective prioritisation of the UCAs. As part of the approach to better communicate the prioritisation results and plan the next steps of system development, a dynamic-scaling prioritisation matrix was developed to capture different sets of prioritised UCAs. The approach was applied to a real project to improve the safe operations of Electric Vertical Take-off and Landing (eVTOL). The results highlighted critical UCAs that need to be prioritised for safer eVTOL operation. 318 UCAs were identified in total. Based on the application of the prioritisation methodology, 110 were recognized as high-priority UCAs to strengthen the system design.

\end{abstract}

\begin{keyword}
STPA \sep UCAs \sep Safety Management \sep Monte Carlo Simulation \sep Prioritization Matrix \sep Expert Judgement \sep Severity Impact Factor.

\end{keyword}

\end{frontmatter}

\section{Introduction}
With the rapid advancement of technology, the complexities of the technologies, as well as the processes of managing these technologies, have also significantly increased. Identifying potential safety flaws due to unexpected emergent behaviors of the process needs to be prioritised at the early phases of the development. System-theoretic Process Analysis (STPA), which is part of the System-Theoretic Accident Model and Processes (STAMP), has been a promising safety analysis method that can be applied in the early phases of system development. STPA has been used in various studies across different domains. Including aviation \cite{allison2017systems,schmid2018laser}, medical sector~\cite{wong2020stamping,yamaguchi2019system,chen2021analyzing}, maritime~\cite{yang2020systems,ventikos2020systems}, automotive~\cite{zhou2019hazard,khastgir2021systems,khastgir2017introducing,chen2020identifying}, railway safety management~\cite{tonk2024application}, and the safety analysis in the area of Large Language Models (LLM)~\cite{qi2025safety}. Especially in the aviation domain, STPA was highly recognized as a recommended methodology in the report \cite{thomas2024evaluation}. The report also showed findings of a collaborative effort by subject matter experts (SME) from the UK Civil Aviation Authorities (CAA) to study STPA and assess its relevance to aviation safety, including safety management, aircraft development, safety assessment, and certification. It also showed how STPA can effectively identify real design flaws and address the gaps. Unlike traditional safety analysis methods such as FTA, FMEA, and HAZOP, STPA goes beyond these approaches by detecting software-related and non-failure scenarios that these traditional methods may overlook
\cite{james2023comparison,merrett2019comparison,sun2022comparison}. 

A standard STPA starts with identifying potential events that need to be prevented, termed as Losses (i.e., Step 1). It then models the interactions of the system stakeholders, which is termed Control Structure (i.e., Step 2). In the next step, any possible unsafe behaviours of each system stakeholder (extracted from the Control Structure) under various circumstances are identified, which is called Unsafe Control Actions (UCAs) in Step 3. Each UCA is then further analysed to understand its causal factors by reviewing the control loop of the controller and its interactions with other system components. STPA can identify a diverse set of causal factors, including but not limited to communication errors, flawed control algorithms, flawed processes, conflicted controls, and missing or inadequate feedback. 

\subsection{Research Gaps and Motivations}
STPA has been a promising approach to identify and address the safety concerns of a complex system. However, this means that the analysis would lead to a massive number of results, which would then require a significant amount of time and effort to manage. There has been a growing concern about the large volume of analytical outputs (i.e., UCAs and requirements) and the challenge of managing them effectively without compromising their granularity, completeness, and manageability. Since all the UCAs identified from STPA are relevant and need to be considered further, overlooking any UCA may raise critical questions about the accuracy and completeness of the results. The UCAs need to be prioritised so that the corresponding requirements can be identified and addressed first to prevent or mitigate these most critical UCAs. However, prioritising the UCAs has become a common challenge as the ranking of the UCAs can be subjective to different system stakeholders. Even if the UCAs are prioritised, there are also challenges of communicating the list of prioritised UCAs with other stakeholders due to a lack of visibility of the contributing factors for the ranking. Therefore, the motivation of this work is to resolve the following research gaps:

\begin{itemize}
    \item RQ1: How can we better utilize STPA results to strategically improve the system development process?
    \item RQ2: How can we objectively prioritise all the UCAs from STPA?
    \item RQ3: How can the prioritised UCAs be better communicated with the stakeholders?
\end{itemize}

\subsection{Paper contribution and Novelty}
In this paper, the authors aim to present a new methodology that combines the STPA methodology and the prioritisation concept to manage a significant number of UCAs identified from a complex system. The approach maintains the ultimate analysis goal, which is to effectively eliminate identified risks across various causal factors. By considering various criteria such as severity, impact factor, and expert judgment input, it allows an effective prioritisation of UCAs and ensures that critical ones are addressed immediately. In the absence of a scientifically definitive approach for prioritisation of STPA results, empirical validation becomes crucial. To demonstrate the effectiveness of the proposed model, a case study from the aviation sector was conducted and will be presented in a later section.

To identify the criteria of an ideal prioritisation framework. A study was conducted to explore existing surveys of prioritisation theory \cite{saaty1994fundamentals}. Table \ref{Tab:prioritisation approach requirements} illustrates the ideal criteria of a good prioritisation approach (left column) and how our proposed approach can fulfil these criteria (right column): 

\begin{table}[tb]
\centering
\caption{How our approach meets the criteria of an ideal prioritisation framework}
\label{Tab:prioritisation approach requirements}
\begin{tabularx}{\linewidth}{XX}
\toprule
\textbf{Criteria of an Ideal Prioritisation Framework} & \textbf{Contribution of our Approach to the Criteria} \\
\midrule
Structure problem as a hierarchy & Using STPA step 2 to derive the CIF.\\
 \midrule 
Involve domain expert's inputs & Both inputs from STPA analysis and stakeholders\\  \midrule 
Represent those inputs into significant numbers & Use the Simple Additive Weighting\\ \midrule 
Analyse the sensitivity to variety in judgment & MCS \\ \midrule 
Synthesize these results for a better visualisation of the findings & Prioritisation Matrix\\  

\bottomrule
\end{tabularx}
\end{table}

\begin{figure*}[tb]
    \centering
    \includegraphics[width=0.8\linewidth]{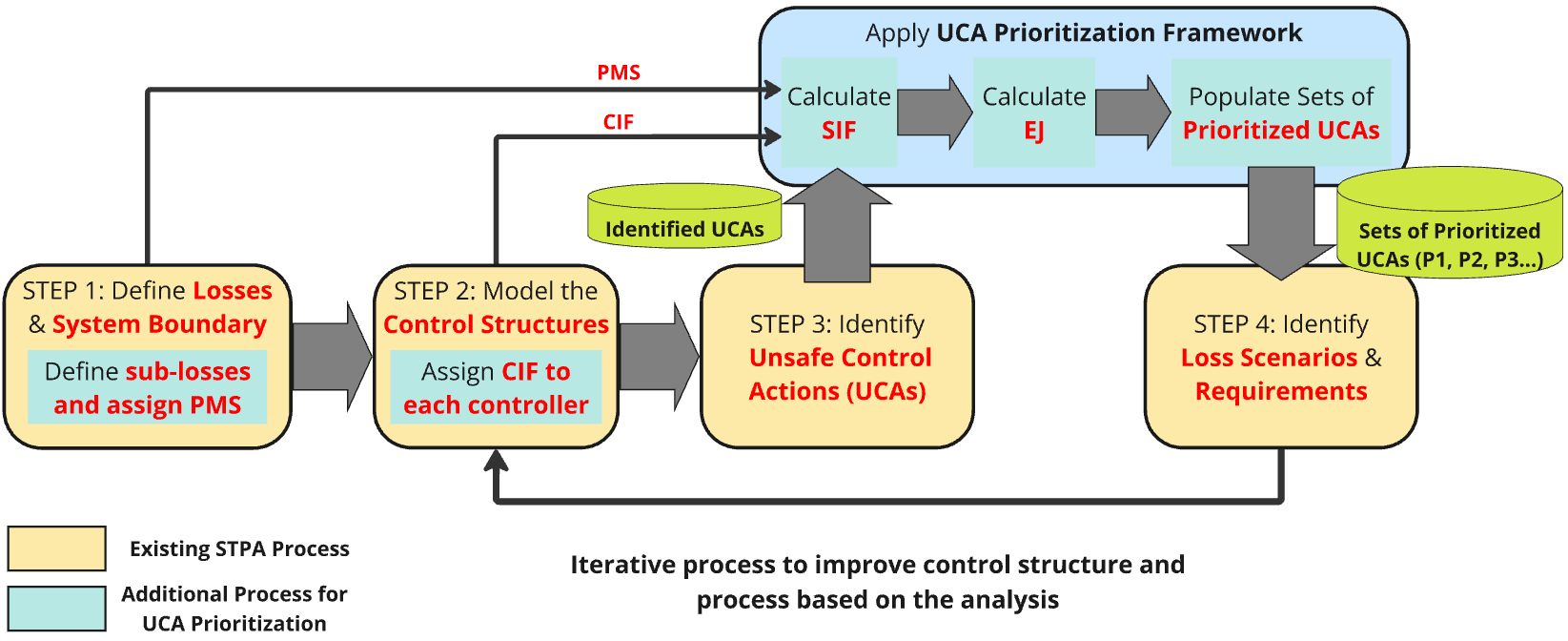}
    \caption{Flowchart of integrating the STPA and the Prioritisation concept}
    \label{fig:Flowchart}
\end{figure*}

In this paper, a new approach to address the research gaps mentioned above is introduced. In the first stage, we will follow the standard STPA steps outlined by Professor Nancy Leveson  \cite{leveson2018stpa} to define Losses, model the Control Structure, and identify UCAs. Before the next step of the standard STPA (i.e., identifying loss scenarios of the UCAs), our approach will be applied to prioritise these identified UCAs. Figure \ref{fig:Flowchart} illustrates the standard STPA process (blocks filled in yellow) and the additional steps called UCA Prioritisation Framework (blocks filled in blue).

The UCA Prioritisation Framework utilizes the outputs from both standard STPA steps 1 and 2, called Pre-Mitigation Severity (PMS) and Controller Impact Factor (CIF). Both PMS and CIF are identified based on STPA analysts and are needed to enable the calculation of the Severity Impact Factor (SIF), which forms part of the axis for the final UCA Prioritisation Matrix. The second part of the prioritisation framework includes Expert Judgement (EJ). In this step, each UCA is reviewed and scored by the relevant domain experts to evaluate its criticality, controllability, and detectability. The sensitivity of changes to the scores are then analysed and addressed using the Monte Carlo Simulation (MCS). Both SIF and EJ are then summarized and captured in the UCA Prioritisation Matrix for better visibility and communication of the results. The framework will be elaborated in the Methodology section.

The key contributions of this work are listed below:
\begin{enumerate}
    \item The decision-making process and operation and organisational STPA analysis are improved in terms of effectively shortening the time needed to mitigate catastrophic losses.
    \item UCAs are prioritised based on objective criteria that incorporate contributions from both domain experts and safety analysts.
    \item High-priority UCAs are primely mitigated to enhance the robustness and reliability of the system design.
    \item The visualisation of the results is improved to enable better communication during development, using dynamic scaling of the data in the Prioritisation Matrix. 
\end{enumerate}

\section{Literature Review }
\subsection{Prioritisation in STPA}

The concept of setting priorities for decision-making in STPA is sought due to the speed of growth nowadays in complex systems that involve more caution about operational safety, as discussed in many studies \cite{zikrullah2018prioritization,blindheim2023risk}. 

A study presented in \cite{gil2019toward} examines expert judgments and incident reports to prioritise safety control actions. The study begins by identifying control actions, followed by an analysis of their impact on collisions. Finally, it explores the relationship between expert judgments and causal factors. This investigation underscores the critical role of control actions in accident prevention. However, the method relies on historical data, including past accident reports and expert evaluations, to assess the effectiveness of control measures.

An application of the Risk Priority Number (RPN) is presented in \cite{kim2021utilization}. It proposes a method for managing a large number of UCAs and loss scenarios, emphasizing the importance of prioritising UCAs to concentrate on design aspects with lower risk. However, as the operating context may change, further research is required to refine the assignment of risk priority numbers.

A methodology for prioritising risk control measures in connected automated vehicles (CAVs) is presented in \cite{liu2024quantitative}. It proposes an improvement to the traditional STPA by integrating additional frameworks, such as sensitivity analysis and propagation techniques within Bayesian networks. This integration enables a quantitative prioritisation of UCAs, emphasizing those that significantly affect overall system reliability. However, the subjectivity and the effectiveness of probabilistic data and expert judgment may introduce work limitations that need to be addressed.

Integrating STPA with Operational Design Domain (ODD) to extract and evaluate ODD and relevant loss scenarios was explored in \cite{nakashima2025addressing}. This approach was demonstrated on a Japanese autonomous container ship. A decomposition of the operational context is used to identify process variables, enabling the capture of discrepancies within the process model. The evaluation was based on several metrics, including risk of each process model, severity, and occurrence, which were used to assess the ODD and define relevant scenarios. A potential limitation of this work is the reliance on subjective judgment in the risk assessment.

The afore-mentioned studies present methodologies to manage a large number of UCAs and loss scenarios. The importance of prioritisation is highlighted and needs to be considered to focus on the most critical design aspects with high risk. However, many of these proposed studies lack objectivity or need more automated processes to facilitate implementation.

\subsection{Methodologies for Reducing Uncertainties}

In the face of the rapid increase in the complexity of decision-making, especially when safety is a subject of disagreement, the insight of stakeholders is required to participate and share their perceptions of the problem. That’s why it is highly recommended to combine the valuable insights from domain experts with existing models that allow for more objective results, preventing safety from being compromised by uncertainty.

In modern engineering, mathematical tools for sensitivity analysis are widely used. These tools can be classified into two categories:
\begin{itemize}
    \item Deterministic: A deterministic model has no randomness. Given the same initial conditions, it will always produce the same results.
    \item Probabilistic: A probabilistic model includes randomness. As a result, even with the same initial conditions, its outcomes may vary with each execution.
\end{itemize}

Safety is far from being an exact science. We cannot try to predict all risks, but we can minimize them, therefore advocating for MCS \cite{Pidd2023}. MCS is a technique invented during the Manhattan Project (us nuclear bomb development) that uses repeated random sampling to estimate the properties of complex systems. 
It is mainly used in finances \cite{uzoh2021development,fu1995sensitivity,Boyle1977}, biomedical and healthcare \cite{Rose2023}, and environmental modelling\cite{Pianosi2015}. In \cite{Pidd2009}, MCS is used as an essential method for exploring the subsequent impact of uncertainty on management science modelling. Pidd spots how the MCS allows the quantification of the variability in outcomes to improve robustness.

In the context of decision analysis, in scenarios where expert judgements are subject to uncertainty, MCS enables the visualisation of the distributions after running random simulations to identify sensitivity to changes. The importance of probabilistic thinking using MCS in decision-making is highlighted in \citep{GoodwinWright2014}.

Moreover, Monte Carlo simulations can incorporate sensitivity analyses to identify which expert inputs most significantly influence the overall ranking or outputs, as well as to determine which criteria remain stable under variations. Some argue that the simulation itself serves as a comprehensive sensitivity analysis, thereby eliminating the need for a separate one. However, if there is uncertainty about the elicited probability distributions or the model’s structure, it is advisable to examine how changes in these assumptions affect the simulation results. If such modifications produce only minor effects, the original model may be considered adequate \cite{GoodwinWright2014}.

Clause B.25 of the risk-management standard ISO 31010 \cite{ISO31000:2018} recommends Monte Carlo as one technique for assessing uncertainty: “Monte Carlo provides a means of evaluating the effect of uncertainty on systems.”

\subsection{Risk Matrix}

A clear and effective visualization of results is crucial, especially in scenarios where risk assessment is necessary due to organizational or visualization-related factors. A proper representation of data ensures that risks are accurately communicated and understood, facilitating better decision-making.

As previously mentioned, safety is not an exact science, and ranking results based solely on unique numerical values can be misleading, raising significant concerns regarding accuracy and reliability.

Risk matrices have been the most widespread risk management tools in use today. The matrices with color-coded ranking are inherently simple to understand. According to \cite{riskmatrixreview}, risk matrices promote robust discussions. It offers some consistency in prioritising risks and helps decision-makers to focus on the highest priority risks. The use of Risk Matrix for prioritisation and assessing risks has been recommended in different standards, including ISO 31000 (Risk Management) \cite{ISO31000:2018} and across different fields in Healthcare \cite{lemmens2022}, engineering and project management \cite{bao2022}, and in aviation safety \cite{gray2019}. 

However, traditional risk matrices also have their weaknesses. \cite{nancyriskmatrix} summarizes some of the most glaring weaknesses of the traditional risk matrices. These include a) Lack of Granularity, b) Inaccurate Quantitative Analysis, and c) General Heuristic Biases. This highlights the need to have an improved version of the risk matrix. To address this issue, we have adopted an approach that emulates the widely used Risk Assessment Matrix, incorporating its principles into our framework. This adapted methodology follows established guidelines to construct a new decision-support tool, which we refer to in our paper as the Prioritisation Matrix (P-Matrix).

\section{Methodology}
This section introduces a framework to enable an objective prioritization of UCAs. The framework mainly consists of two parameters called Severity-Impact Factor (SIF) and Expert Judgement (EJ). The inputs required to identify SIF and EJ are collected by STPA analysts and domain experts separately, which are introduced in the following subsections.

\begin{figure}[tb]
    \centering
     \includegraphics[width=0.9\linewidth]{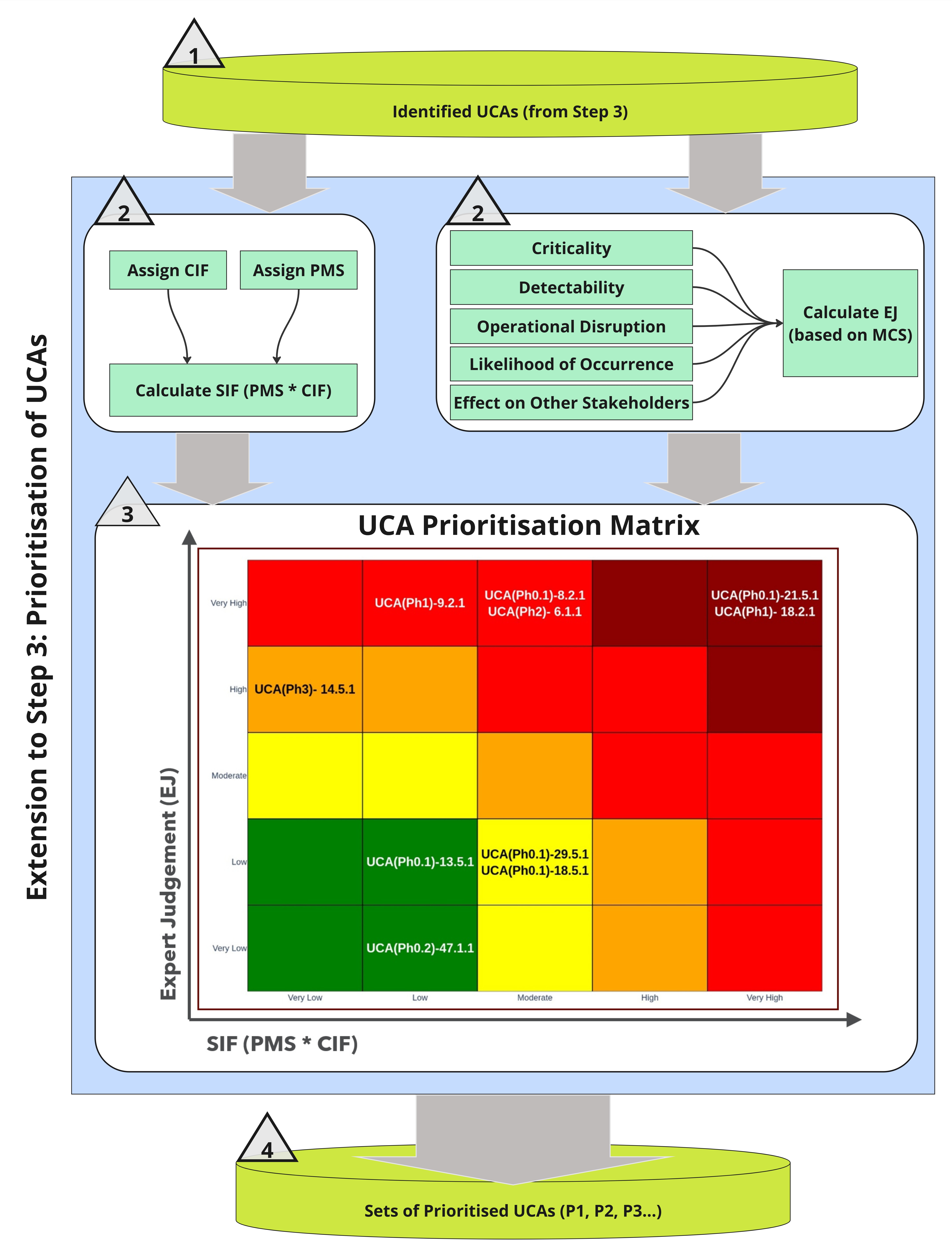}
    \caption{Flowchart of the proposed UCA Prioritisation Framework}
    \label{fig:metho}
\end{figure}

\begin{table}[tb]
\label{tab:TABLE 1} 
\caption{Losses Identified following Traditional STPA Approach}
\centering
\begin{tabular}{|>{\centering\arraybackslash}p{0.1\linewidth} |p{0.8\linewidth}|} 
\hline
\textbf{ID} & \textbf{Description}\\
\hline
\textbf L1& Human Loss: Loss of life or injury to 1st (eVTOL crew), 2nd (passengers), or 3rd parties (anyone outside eVTOL). \\
\hline
\textbf L2 & Material Loss: loss of or damage to the eVTOL or surrounding item/property/infrastructure. \\
\hline
\textbf L3 & Mission Loss: Loss of transportation mission. \\
\hline
\textbf L4 & Consumer Demands Loss: Loss of customer satisfaction or public confidence in eVTOL. \\
\hline
\textbf L5 & Business Goal Loss: Loss of the business goal of eVTOL Operator. \\
\hline
\end{tabular}
\end{table}

\begin{table}
\centering
\label{tab: TABLE 2}
\caption{Refined Losses for PMS Identification}
\begin{tabular}{|>{\centering\arraybackslash}p{0.03\linewidth}|>{\centering\arraybackslash}p{0.05\linewidth}|p{0.65\linewidth}|>{\centering\arraybackslash}p{0.06\linewidth}|} 
\hline
\multicolumn{2}{|>{\centering\arraybackslash}p{0.08\linewidth}|}{\textbf{ID}} & \textbf{Description} & \textbf{PMS} \\ 
\hline
L1 & L1.1 & \textbf{Catastrophic:} Total fatalities (loss of life) & 20 \\ \hline 
 & L1.2 & \textbf{Hazardous:} Multiple Fatalities and/or serious injuries that include mental health losses (e.g., trauma). & 19 \\ \hline  
 & L1.3 & \textbf{Major:} Multiple serious injuries and/or multiple injuries & 18 \\ \hline  
 & L1.4 & \textbf{Minor:} Minor/non-serious injuries. & 10 \\ 
\hline
L2& L2.1 & \textbf{Catastrophic:} Complete loss of the aircraft (loss of up to 100\% of cost). & 17 \\ \hline  
 & L2.2 & \textbf{Hazardous:} Serious or fatal damage to the material (loss of up to 75\% of cost). & 13 \\ \hline  
 & L2.3 & \textbf{Major:} Major damage to the material (loss of up to 50\% of cost). & 8 \\ \hline  
 & L2.4 & \textbf{Minor:} Minor damage to the material (loss of up to 25\% of cost). & 5 \\ 
\hline
L3& L3.1 & \textbf{Catastrophic:} Complete of the tactical mission that significantly affects the strategic mission. & 15 \\ \hline  
 & L3.2 & \textbf{Hazardous:} Complete loss of the tactical mission. & 9 \\ \hline  
 & L3.3 & \textbf{Major:} Partial loss of the tactical mission. & 4 \\ \hline  
 & L3.4 & \textbf{Minor:} Minor degradations of the tactical mission. & 1 \\ 
\hline
L4& L4.1 & \textbf{Catastrophic:} Complete loss of customer satisfaction/consumer demands(up to 100\%). & 16 \\ \hline  
 & L4.2 & \textbf{Hazardous:} Serious loss of customer satisfactions/consumer demands(up to 75\%). & 12 \\ \hline  
 & L4.3 & \textbf{Major:} Major loss of customer satisfactions/consumer demands(up to 50\%). & 7 \\ \hline  
 & L4.4 & \textbf{Minor:} Minor loss of customer satisfaction/consumer demands(up to 25\%). & 2 \\ 
\hline
L5& L5.1 & \textbf{Catastrophic:} Complete loss of business goals (up to 100\% of total business goals). & 14 \\ \hline  
 & L5.2 & \textbf{Hazardous:} Serious loss of business goals (up to 75\% of total business goals). & 11 \\ \hline  
 & L5.3 & \textbf{Major:} Major loss of business goals (up to 50\% of total business goals). & 6 \\ 
 \hline 
 & L5.4 & \textbf{Minor:} Minor loss of business goals (up to 25\% of total business goals). & 3 \\ 
\hline
\end{tabular}
\end{table}

\subsection{Severity-Impact Factor (SIF)}

SIF mainly consists of two parameters, called Pre-Mitigation Severity (PMS) and Controller Impact Factor (CIF).

\subsubsection{Pre-Mitigation Severity (PMS)}

PMS defines the severity of a risk before any mitigation is implemented. In this case, it defines the severity of each UCA. Based on the traditional process of STPA, each UCA should lead to at least one system-level hazard or loss that is identified in Step 1, as otherwise, the control action would be safe. To enable the identification of the PMS of each UCA, we extended the Step 1 results by further refining each loss based on Design Assurance Levels (DAL): 1) Catastrophic; 2) Hazardous; 3) Major;  4) Minor; and 5) No Effect). These guidewords define the severity levels, with `Catastrophic' representing the highest severity and incrementally `No Effect' representing the lowest severity (i.e., zero severity). They have been commonly used in many standards such as DO-178C (Software Considerations in Airborne Systems and Equipment Certification) \cite{sae2010arp4754a}. In this work, `No Effect' is not considered.

Table \ref{tab:TABLE 1} shows the original losses identified by system stakeholders. It is important to note that the losses identified following STPA are not only limited to safety-critical losses such as L1 (Human Loss) but also non-safety-critical losses like L2-L5. Because they are all unacceptable to the stakeholders. Table \ref{tab: TABLE 2} shows the refined version of the losses identified in Table \ref{tab:TABLE 1}. Each loss (i.e., L1-L5) has been refined according to DAL. For example, the loss L1 (Human Loss) was refined based on DAL, representing four levels of severity for the human loss. L1.1 (Catastrophic: Total fatalities) indicates the most severe level for human loss - i.e., loss of human life. L1.2 is slightly less severe compared to L1.1, which indicates life-long physical injuries or mental health loss. L1.3, which is slightly severe compared to L1.2, indicates serious injuries. Lastly, L1.4, which is the least severe sub-loss of L1, indicates minor injuries. There are in total twenty sub-losses, each of which has a unique PMS value assigned, ranging from 20 (most severe) to 1 (least severe). To finalize the ranking of PMS values, a series of workshops was conducted that involved the stakeholders of the systems, which were also the stakeholders of the project. This includes stakeholders from the areas of the regulator (UK Civil Aviation Authority), air traffic service provider (NATS), eVTOL Operator (Lillium), and vertiport and infrastructure management service provider (Skyport). Each stakeholder provided their proposed ranking of the PMS. These rankings were then further jointly reviewed and analyzed to create a final list of PMS rankings. This is illustrated in the last column of Table \ref{tab: TABLE 2}. Of the twenty sub-losses, L1.1 (PMS 20), L1.2 (PMS 19), and L1.3 (PMS 18), representing three levels of human losses, have the highest PMS value. This is then followed by L2.1 (PMS 17), which indicates the complete loss of the aircraft and consequently, L4.1 (PMS 16), which indicates the complete loss of consumer demands.

Whilst in traditional STPA, each identified UCA links to at least one loss. For example, considering the: `Licensed Aerodrome provides a \textbf{Hold} command too long when the eVTOL is in the air, approaching the landing pad, and has a very limited battery range left, the aircraft, which is powered solely by battery, could, in the worst-case scenario, run out of battery and fall off. This would lead to both L1 (Loss of Life) and L2 (Loss of aircraft). As an updated process, for the same UCA, rather than linking to the L1 and L2, each identified UCA links to one of the sub-losses of L1 and L2. In this case, the UCA links to L1.1 and L2.1. Because aircraft falling could, in the worst-case scenario, lead to loss of human life and complete loss of the aircraft. It is important to note that for a UCA that is linked to multiple sub-losses, the sub-loss with the highest PMS value is assigned to the UCA. Therefore, the example UCA has a PMS value of 20.

\subsubsection{Controller Impact Factor (CIF)}
\begin{figure*}[p]
    \centering
    \includegraphics[width=0.95\textheight,height= 1.5\linewidth,keepaspectratio,angle=90]{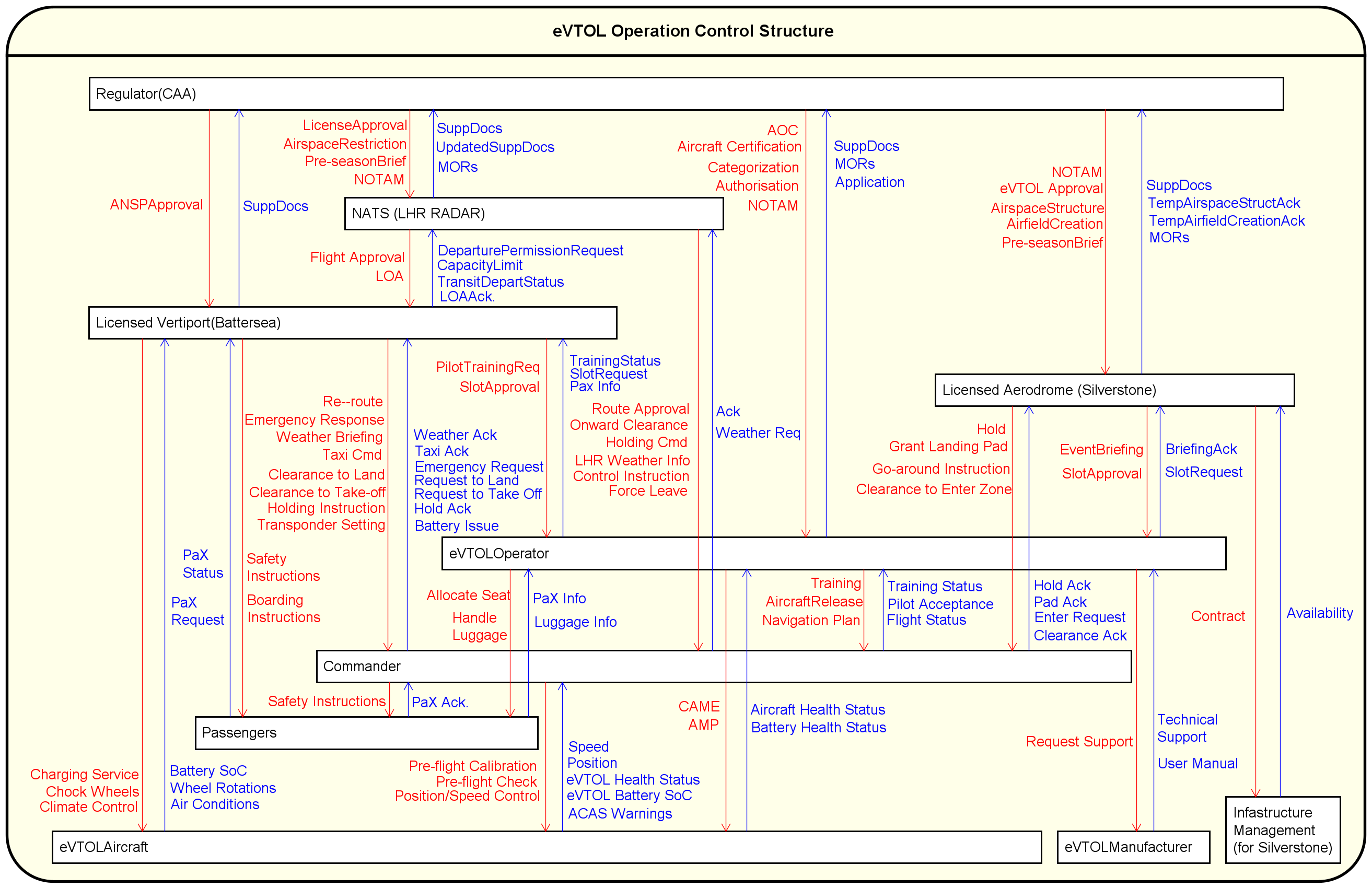}
    \caption{Control Structure of eVTOL Operation}
    \label{fig: control structure}
\end{figure*}

CIF is a parameter used to quantify a controller's impact or influence.  In STPA, a control structure (created in Step 2) fundamentally consists of a group of hierarchically distributed nested control loops. Each control loop is made up of a controller, a CA that the controller sends, a controlled process that receives and implements the CA, and the feedback from the controlled process to enable the controller to update the CA if necessary. A controller in the context of a socio-technical system is not necessarily just a physical or digital controller, but can also be a decision-maker as part of an organization or department. Figure \ref{fig: control structure} illustrates the control structure of a generic eVTOL operation. A diverse set of stakeholders are involved in the system of eVTOL operation, this includes: a) the regulator that reviews and approves all flight applications, infrastructure management, aircraft certification, and commander training; b) the air navigation service provider; c) the licensed vertiport or aerodrome that maintains the infrastructure and provides training to the commander for the specific operation; d) the eVTOL operator that owns and maintains the aircraft; and e) the commander that directly engages with the aircraft during the flight. Each block has a different role depending on where it is in the control loop. For example, considering the block [NATS (LHR RADAR)] in Figure \ref{fig: control structure}, when it is implementing the CAs from [Regulator (CAA)], it acts as a controlled process within the control loop between [Regulator (CAA)] and [NATS (LHR RADAR)]. When it is sending the CA to [Licensed Vertiport (Battersea)], it acts as a controller within the control loop between [NATS (LHR RADAR)] and [Licensed Vertiport (Battersea)]. There are blocks in the control structure that are solely controlled processes, located at the bottom of the control structure. This includes [eVTOL Manufacturer], [Infrastructure Management (for Silverstone)], [eVTOL Aircraft], and [Passengers]. Because they do not send any CAs , CIF does not apply to these controlled processes.

To rank the impact of each controller of the control structure, i.e., their CIF value, it is important to understand how many blocks are affected by the decisions (i.e., the CAs) made by that particular controller. In STPA, a control structure is created in such a way that functional blocks are located hierarchically - i.e., controllers are above the controlled processes. We proposed a mechanism to rank the CIF based on the hierarchy of the blocks. Considering the same control structure in Figure \ref{fig: control structure}, the block [Regulator (CAA)] has the highest CIF value because it is located at the top level of the control structure. The behaviors of and decisions made by the other blocks can directly or indirectly be affected by the CAs from [Regulator (CAA)]. The next highest CIF value is assigned to [NATS (LHR RADAR)], which is located at the second highest level of the control structure. The [Commander] block, which directly engages with the passengers and aircraft, has the lowest CIF in this control structure. While it may sound atypical that a commander has the lowest impact factor here, it is important to note that the ranking of CIF is closely linked to the size of the analysed system. In this work, the scope of the analysis is to understand potential issues with the eVTOL operation, which involves regulatory, organizational, and operational management. A UCA from a high CIF controller (like the Regulator) could potentially affect many other stakeholders simultaneously, such as the operation of all the vertiports, eVTOL operators, and commanders of different eVTOL aircraft. It is therefore arguable that the commander, as part of this system, would have a much lower CIF value compared to the regulator.

Table \ref{tab:CIF summary} summarizes the CIF values of all the controllers in the control structure. Once a UCA is identified, the UCA inherits the CIF value of the controller of the UCA. Consider the same UCA in the previous section (i.e., Licensed Aerodrome provides a 'Hold' command too long when the eVTOL is in the air, approaching the landing pad, and has a very limited battery range left) as an example, the UCA is from the controller 'License Aerodrome'. Based on the CIF value in Table \ref{tab:CIF summary}, this UCA would be assigned a CIF value of 3.

\begin{table}
\centering
\caption{\label{tab:CIF summary} Summary of CIF Values for all the Controllers}
\begin{tabular}{|l |c|} \hline
\textbf{Controllers}& \textbf{CIF} \\ \hline 
Regulator (CAA) & 6 \\ \hline 
NATS (LHR RADAR) & 5 \\ \hline 
Licensed Vertiport (Battersea) & 4 \\ \hline 
Licensed Aerodrome (Silverstone) & 3 \\ \hline 
eVTOL Operator & 2 \\ \hline 
Commander & 1 \\ \hline
\end{tabular}
\end{table}

\subsubsection{SIF}
Once both PMS and CIF values are assigned to all the UCAs, the SIF value can then be calculated. SIF is defined as the product of PMS and CIF as in equation (\ref{eq1}). The SIF value will be used together with the Expert Judgement for the final UCA prioritization, which will be introduced in the next section.

\begin{equation}
\label{eq1}
    SIF = PMS * CIF
\end{equation}

\subsection{Expert Judgment}
\label{EJ}

This chapter summarizes the third factor, along with the SIF, which is provided by the STPA analyst. Relying solely on STPA analyst inputs might not be reliable, acknowledging the analysts' lack of domain-specific knowledge.

To efficiently rank the UCAs, this factor primarily involves scoring by domain experts, who have a better understanding of domain-specific risks that might be overlooked by the STPA analyst. Throughout this paper, this factor will be referred to as EJ.

\begin{figure*}[p]
    \centering
     \includegraphics[width=\linewidth]{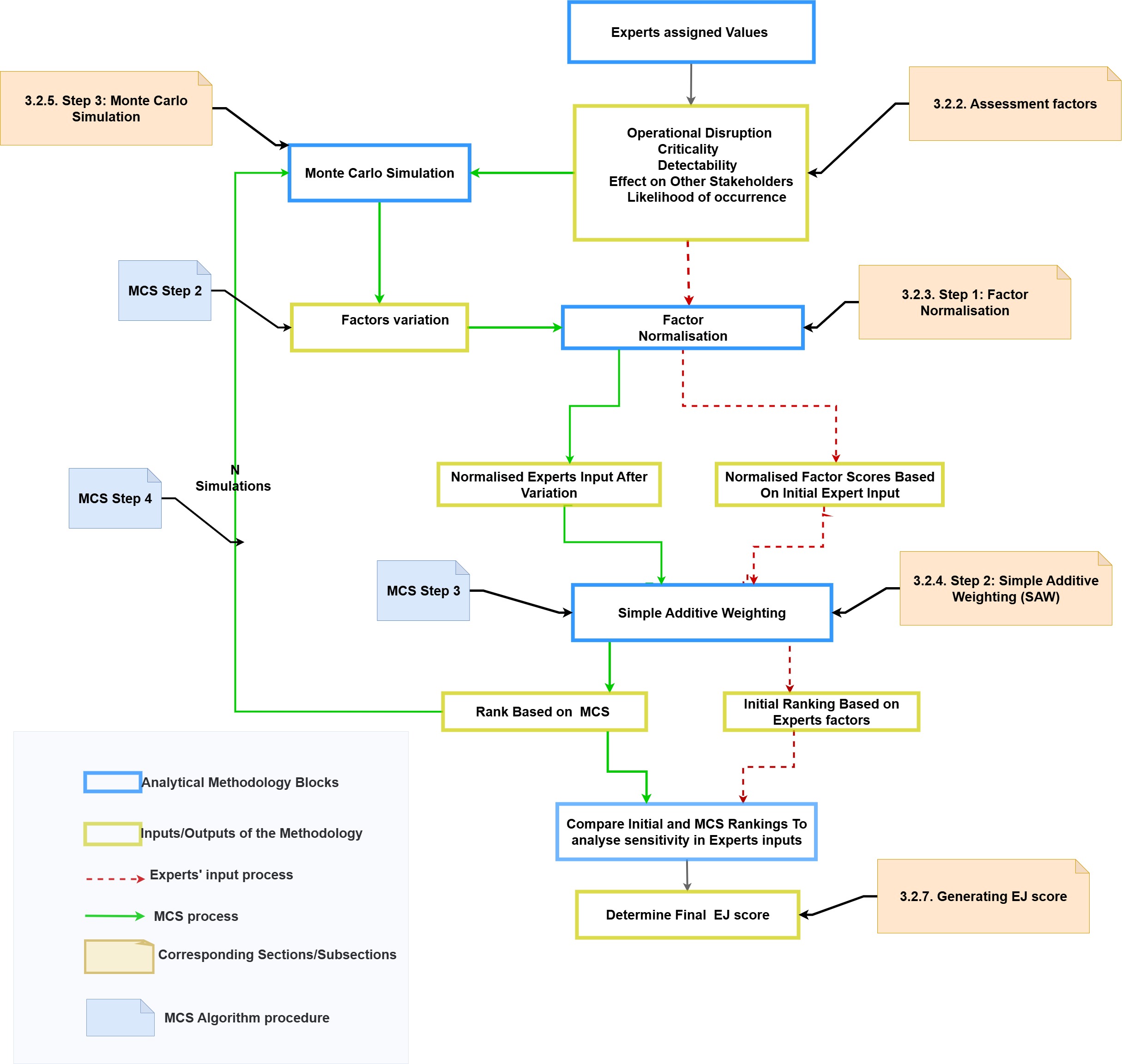}
    \caption{EJ score Calculation Process, with each step annotated with the corresponding section/subsection where it is described in detail}
    \label{fig:EJ score Calculation Process}
\end{figure*}
\subsubsection{Criteria Definition}  

In the process of prioritising UCAs, domain experts are consulted mainly to set priorities for decision-making in different domains. Thus, their ability to predict hazards is based on knowledge and experience to interpret data from prior experience. However, their perspective of viewing risk is still questionable, and their assessment remains subjective, which might change over time or based on their understanding. The chosen criteria also play a crucial role in expert judgment. The criteria must be based on different aspects to cover all perspectives when a risk arises.
Relying solely on domain experts in such analyses, where risk prevention is indisputable, highlights the need to reduce subjectivity and uncertainty in assessments. In the context where this methodology was applied, contributions from domain experts was essential due to their specialized knowledge of the aviation industry. This was particularly crucial in cases where STPA outcomes impacted regulatory changes, as their expertise was vital in defining appropriate safety barriers.
At the core of the problems that our method addresses is making the expert's assessments objective. Thus, this approach requires: 
\begin{itemize}
    \item Involvement of stakeholders' input;
    \item Selection of key criteria;
    \item Analyses of the sensitivity to changes in judgment.
\end{itemize}

\subsubsection{Assessment factors}
Decision-making relies essentially on the criteria chosen to be evaluated by the domain experts and the point-scoring method enacted.
\subsection*{Key criteria}
In traditional expert judgment, severity, controllability, and likelihood  are  the key criteria to assess risk. 
For a complex system with innovations in its design, and inclusion of a diversified set of stakeholders, it becomes particularly challenging to adopt traditional criteria for risk assessment. One major issue is the absence of past data, making it difficult to evaluate risks using conventional approaches. As a result, these criteria are no longer valid for such systems.  

This necessitates the development of new criteria that align with the goal of the analysis and enable a comprehensive evaluation of UCAs.  
Multi-criteria decision-making (MCDM) is a tool used to set priorities by evaluating different options based on multiple criteria, each of which may have a different level of importance or rating \cite{saaty1994fundamentals}.
Enacting the MCDM in our approach, we will refer to the composition of UCAs to achieve this. A UCA statement consists of the CA, the controller, and the context, the combination of the actor (i.e., the controller) with the behaviour (i.e., the CA) in the particular circumstance (i.e., the context) potentially leads to losses. From each component of the UCA composition, we will derive a corresponding factor to ensure a complete assessment.
Also, these criteria need to be tailored to the domain expert.
\begin{itemize}
    \item \textbf{Operational Disruption}: Whether these UCAs can lead to operational disruptions when control actions do not behave as intended.
    \item \textbf{Criticality}: Whether this UCA is critical, looking at the severity in stakeholder's point of view, Criticality from experts differs  from the PMS.
    \item \textbf{Detectability}: Refers to how easily a UCA can be detected before it leads to a loss.
    \item \textbf{Effect on Other Stakeholders}: The occurrence of UCAs can impact one or multiple stakeholders.
    \item \textbf{Likelihood of Occurrence}: Refers to whether this UCA has not been mitigated by pre-existing regulations and is likely to occur or not. This factor is not related to the actual probability used in traditional risk assessment based on historical data. Instead, it is determined by whether this UCA is mitigated by existing regulations and its likelihood of occurrence.
    
\end{itemize}
This classification has the advantage that each criterion is more or less independent of the others, while complementing one another. Additionally, each criterion addresses a specific question that helps experts to visualise the UCA and its controlling factors, for which the domain expert can evaluate and provide relative scores.

\subsection*{Turning criterion into Priorities}
A potential next step is to allocate an intensity to each criterion; these intensities should guide the experts in their assessment and calculate an overall priority.
These intensities must be clear and unique to each criterion \cite{GoodwinWright2014}. The application of this methodology is in the context of eVTOL operation. Thus, these criteria need to be adapted to this context as shown in the table  \ref{Tab:Criterions and intensities}. 

\begin{table*}
\caption{Criteria for setting priorities and their intensities}
\label{Tab:Criterions and intensities}
\begin{tabularx}{\linewidth}{|p{3cm}|X|}
    \hline
    \textbf{Criterion} & \textbf{Intensities} \\ 
    \hline

    Operational Disruption & 
    \begin{itemize}
        \item \textbf{High Impact}: The UCA leads to severe disruptions, including system-wide airspace management breakdowns, multiple flight cancellations, or long-term restrictions.
        \item \textbf{Medium Impact}: The UCA results in localized disruptions that impact only a subset of operations.
        \item \textbf{Low Impact}: The UCA causes very limited disruptions with manageable consequences. 
    \end{itemize} 
    \\ \hline

    Criticality & 
    \begin{itemize}
        \item \textbf{High Risk}: A UCA causes an immediate risk to eVTOL safety or operations, leading to catastrophic consequences.
        \item \textbf{Moderate Risk}: A UCA causes operational delays or significant safety concerns but allows time for corrective action.
        \item \textbf{Low Risk}: A UCA has minimal impact, such as causing minor delays or requiring rerouting, without posing a significant risk to eVTOL operations or safety. 
    \end{itemize} 
    \\ \hline

    Detectability & 
    \begin{itemize}
        \item \textbf{Low Detectability}: The UCA is inherently difficult to detect due to limited monitoring capabilities or delayed feedback, leading to potential long-term risks before discovery.
        \item \textbf{Moderate Detectability}: The UCA can be identified but requires significant effort, manual intervention, or time to detect and correct, often delaying the response.
        \item \textbf{High Detectability}: The UCA is easily and promptly detected, either through automated systems or real-time monitoring, allowing for rapid intervention with minimal disruption.
    \end{itemize} 
    \\ \hline

    Effect on Other Stakeholders & 
    \begin{itemize}
        \item \textbf{Significant Impact}: The UCA affects multiple stakeholders, causing a breakdown in communication, coordination, or responsibilities.
        \item \textbf{Moderate Impact}: The UCA affects only some stakeholders and can be managed through coordination.
        \item \textbf{Minimal Impact}: The UCA has little to no impact on stakeholders or causes minimal inconvenience.
    \end{itemize}  
    \\ \hline

    Likelihood of Occurrence & 
    \begin{itemize}
        \item \textbf{1}: Not mitigated by pre-existing regulations and likely to occur.
        \item \textbf{0}: Mitigated by pre-existing regulations and unlikely to occur.
    \end{itemize}
    \\ \hline
\end{tabularx}
\end{table*}

The next step to finalize this process is to assign a score to each intensity, ranging from 1 to 3, based on its priority. After identifying the key criteria of expert judgment, we will follow the methodology depicted in the figure \ref{fig:metho} and each step will be detailed in the following subsections. It is important to note that to simplify the notation in this paper, each criterion will be called $f$.



\subsubsection{Step 1: Factor Normalisation}
This step ensures that each factor contributes equally to the ranking, regardless of its original intensity. 
 Min-max normalisation following the equation \ref{nor} is commonly used in MCDM to avoid any bias caused by differing scales of factors. This normalisation is crucial when employing methods like SAW (Simple Additive Weighting) to ensure fair weighting \cite{cahyapratama2018}.

\begin{equation}
    \label{nor}
    f_{\text{norm}} = \frac{f - \min(f)}{\max(f) - \min(f)}
\end{equation}

The example presented in \cref{tab:example_data} illustrates the application of the \cref{nor}. In this example, $f$ presents the criticality obtained from the dataset presented in \cref{tab:example_data}, which are $\min(f)=2$, $\max(f)=3$. Application of the formula on the first UCA (i.e., UCA-1.1.1) is shown below and the factors are depicted in table \ref{tab:Normalisation} after normalization.

\begin{table*}[tb]
\caption{Example dataset of UCA.}
\label{tab:example_data}
\centering
\begin{tabular}{ccccc>{\centering\arraybackslash}p{2cm}}
\toprule
\textbf{UCA} & \textbf{Operational Disruption} & \textbf{Criticality} & \textbf{Detectability} & \textbf{Effect on Other Stakeholders} & \textbf{Likelihood of occurrence} \\
\midrule
UCA-1.1.1 & 3 & 3 & 2 & 3 & 0\\  \midrule 
UCA-1.2.1 & 2 & 2 & 3 & 3 & 1 \\  \midrule 
UCA-2.1.1 & 1 & 2 & 1 & 2 & 1 \\  

\bottomrule
\end{tabular}

\end{table*}

\begin{align*}
f_{\text{C-norm}} &= \frac{f - \min(f)}{\max(f) - \min(f)} \\
f_{\text{C-norm}} &= \frac{3 - 2}{3 - 2}=1 
\end{align*}

\begin{table*}[tb]
\caption{Normalisation of UCA.}
\label{tab:Normalisation}
\centering
\begin{tabular}{lrrrrr}
\toprule
\text{UCA} & {\text{Op. Disruption}} & {\text{Criticality}} & {\text{Detectability}} & {\text{Effect on Stakeholders}} & {\text{Likelihood}} \\
\hline
UCA-1.1.1 & 1.000 & 1.000 & 0.500 & 1.000 & 0.000 \\  \midrule 
UCA-1.2.1 & 0.500 & 0.000 & 1.000 & 1.000 & 1.000 \\ \midrule 
UCA-2.1.1 & 0.000 & 0.000 & 0.000 & 0.000 & 1.000 \\
\bottomrule
\end{tabular}
\end{table*}
\subsubsection{Step 2: Simple Additive Weighting (SAW)}
This step allows these factors to be converted into priorities.
Summing the scores assigned by the expert after the normalisation of each criterion \(f\) as indicated in the equation \ref{SAW}.
This step calculates and aggregates the score for each UCA by summing the normalized values of each factor. Higher scores indicate higher priority.
 
SAW is a recommended application in the MCDM. An initial ranking is provided 
 by summing the normalise factors \cite{cahyapratama2018,saltelli2008,piasecki2019}.
\begin{equation}
\label{SAW}
   S_{\text{UCA}} = \sum_{i=1}^n f_{\text{norm}_i}
\end{equation}
where $f_{\text{norm}_i}$ represents the normalized value of the $i$-th factor for the UCA. As an example, the \emph{Simple Additive Weighting (SAW)} score for each UCA is summarized in \cref{tab:SAW score}.

\begin{table}
    \centering
    \caption{Calculated SAW Score of each UCA}
    \begin{tabular}{ccc}\toprule
         UCA & $S_i$ & Rank\\\midrule
         UCA-1.1.1 &  3.500& 1\\
         UCA-1.2.1 &  3.500& 1\\
         UCA-2.1.1 &  1.000& 3\\ \bottomrule
    \end{tabular}
    \label{tab:SAW score}
\end{table}

\subsubsection{Step 3: Monte Carlo Simulation}

Scoring the factors by the experts is subject to uncertainty. To address this issue, the MCS is needed. 

In this step, MCS is used to reduce relativity and subjectivity in experts' judgment scoring, thereby producing a more reliable assessment and objective results, enhancing decision-making \cite{Pidd2023}.

The main purpose of applying MCS is to explore the 'what if' scenarios. By implementing the model with specific input parameters, analysts can observe how variations in these inputs impact the outputs. A computer typically performs MCS using a specially designed algorithm tailored for this purpose, which include the following steps:
\begin{itemize}
    \item MCS Step 1: Calculate the initial ranking (as described in \cref{SAW}) based on the experts' inputs.
    \item MCS Step 2: Vary each factor by \qty{+-10}{\percent} over 1000 simulations to measure how small input changes affect rankings. The variation of each UCA factor $f$ by \qty{+-10}{\percent}: 
    \[
    f_{\text{sim}} = f \times (1 + \text{random}(-0.1, 0.1))
    \]
    where \(\text{random}(-0.1, 0.1)\) is a uniformly distributed random number between \(-0.1\) and \(0.1\), applying small changes to simulate variability.
    The choice of a \qty{+-10}{\percent} variation in MCS is recommended across different fields to minimize relativity while preserving small variations in each factor’s contribution to the overall ranking.
    
    \item MCS Step 3: Recalculate the SAW (as described \cref{SAW}) using the new factor $f_{\text{sim}}$.
    Recalculation of the SAW score using the modified factor $f_{\text{sim}}$ to obtain a new rank, \(\text{Rank}_{\text{sim}}\), for each factor of UCA.
    \item MCS Step 4: Repeat the trial, computing the Rank Variability.
    \item MCS Step 5: Generating the Overall score of each UCA.
\end{itemize}




    

An example of the application of the entire MCS process is presented below, which include two runs of simulations. Assume the random variations \(X\) are as follows (one for each factor of each UCA):

\subsubsection*{Simulation 1:}
\begin{align*}
 X_{\text{UCA-1.1.1, Op.\ Disr.}}^{(1)} &= -0.10, \\
 X_{\text{UCA-1.1.1, Crit.}}^{(1)} &= +0.05, \\
 X_{\text{UCA-1.1.1, Detect.}}^{(1)} &= -0.10, \\
&\vdots
\end{align*}

(etc.\ for \(\text{UCA-1.2.1}\) and \(\text{UCA-2.1.1}\)). 
After applying each $X_{i,j}^{(1)}$ and recomputing:
\[
S_{\text{UCA-1.1.1}}^{(1)} 
= 3.40,
\quad
S_{\text{UCA-1.2.1}}^{(1)} 
= 3.35,
\quad
S_{\text{UCA-2.1.1}}^{(1)} 
= 1.20.
\]
Hence, the ranks in Simulation~1 remain:
\[
\text{UCA-1.1.1} \;\to\; 1,
\quad
\text{UCA-1.2.1} \;\to\; 2,
\quad
\text{UCA-2.1.1} \;\to\; 3.
\]

\subsubsection*{Simulation 2:}

For the second run, suppose a different set of \(X_{i,j}^{(2)}\). 
The recalculated SAW scores might be:
\[
S_{\text{UCA-1.1.1}}^{(2)} 
= 3.50,
\quad
S_{\text{UCA-1.2.1}}^{(2)} 
= 3.55,
\quad
S_{\text{UCA-2.1.1}}^{(2)} 
= 1.00.
\]
This time, \(\text{UCA-1.2.1}\) edges out \(\text{UCA-1.1.1}\) slightly and takes rank~1, while \(\text{UCA-2.1.1}\) remains rank~3, which implies its uncertainty.

\subsubsection{MCS Algorithm}
This paragraph presents the algorithm used to implement the MCS and to assess how initial ranking is sensitive to changes.

The MCS algorithm was implemented on Google Colab.
The function \texttt{monte\_carlo} performs the MCS and computes the new rankings under different perturbations. We build this function to apply our methodology following the steps: 
\subsubsection*{Function Parameters}
The parameters of the function:\\
\texttt{def monte\_carlo(df, factors, num\_simulations,\\variation\_range)}\\
are initially defined, which are summarized below: 
\begin{itemize}
    \item \textbf{\texttt{df}}: DataFrame that contains experts' inputs.
    \item \textbf{\texttt{factors}}: List of Criteria impacting the rank.
    \item \textbf{\texttt{num\_simulations}}: Number of Monte Carlo iterations.
    \item \textbf{\texttt{variation\_range}}: Percentage variation applied to each factor.
\end{itemize}
\subsubsection*{Function Execution}

        





To execute the function, a loop is firstly executed for 

num\_simulations iterations. Each factor is varied randomly within the specified range by applying the\\ \texttt{function numpy.random.uniform()}. The function variation\_range controls the extent of the variation (e.g., 0.1 for ±10\%). Then, the SAW score is recalculated, and the UCAs are re-ranked based on the new SAW scores. Finally, the rankings are computed using the SAW score, with higher scores receiving better ranks. The results are stored in a pandas DataFrame and returned.

\subsubsection{Generating EJ score}
After generating the Monte Carlo ranking data, additional steps are performed to generate the final rank.

\subsubsection*{Calculating the Average Rank}
The ranking, denoted as \(\text{Rank}_{\text{sim}}\), refers to the ordinal position of each UCA factor after recalculating the SAW score for each simulation. After conducting the MCS, the average rank for each UCA factor is calculated as follows:

\begin{equation}
\text{Average Rank}_{\text{UCA}} = \frac{1}{\text{num\_simulations}} \sum_{\text{sim}=1}^{\text{num\_simulations}} \text{Rank}_{\text{sim}}
\end{equation}
Calculating ${\text{Average Rank}_{\text{UCA}}}$ across multiple simulations provides a measure of central tendency, indicating the position of each UCA in terms of risk priority.

\subsubsection*{Computing Rank Variability}
The standard deviation $\sigma_i$ measures the variability or dispersion of ranks across simulations, reflecting the stability of each UCA's ranking.
The standard deviation of ranks per UCA is computed as:

\begin{equation}
    \sigma_i  = \sqrt{\frac{1}{N} \sum_{i=1}^{N} (\text{Rank}_{\text{sim}} - \text{Average\_Rank})^2}
\end{equation}

\noindent{}where $N$ is \texttt{num\_simulations}.

\subsubsection*{Calculating the Overall Score}

To achieve minimal uncertainty in the final EJ score, the overall score must reflect both \emph{ranking performance} (i.e., how well a UCA is ranked on average) and \emph{sensitivity} (i.e., how much that ranking changes with varying factors). Thus, two complementary metrics are combined.

\subsection*{ EJ Score Central Tendency and Spread}
For each UCA, the MCS produces 

$\text{Rank}_{\text{sim}}$$(\text{sim}=1,\dots,\text{num\_simulations})$.

Their mean and standard deviation are combined as follows:

\begin{equation}
\text{EJ-Score}_{\text{UCA}}=
\text{Average Rank}_{\text{UCA}}+\sigma_{\text{UCA}},
\label{eq:EJscore}
\end{equation}
which summarises the rank distribution produced by the MCS.  
Nevertheless, equation~\eqref{eq:EJscore} does not quantify sampling uncertainty.

\subsubsection*{Confidence-Interval score}

To make uncertainty explicit, the two-sided \qty{95}{\percent} confidence interval (CI) of the mean rank is computed for every UCA:
\begin{equation}
\text{CI}_{95}=
\text{Average Rank}_{\text{UCA}}
\pm
1.96\,
\frac{\sigma_{\text{UCA}}}{\sqrt{\text{num\_simulations}}},
\label{eq:rank_CI}
\end{equation}
where the CI is the long-run proportion of intervals that would contain the true mean rank \cite{mandel2020effect}.
In safety-critical contexts, a conservative approach is required; therefore, the \emph{upper} limit of the interval is used.
\cite{bedford2001probabilistic}:
\begin{equation}
\text{CI}^{\uparrow}_{95}=
\text{Average Rank}_{\text{UCA}}+
1.96\,
\frac{\sigma_{\text{UCA}}}{\sqrt{\text{num\_simulations}}},
\label{eq:upper_CI}
\end{equation}
thereby guaranteeing, with \qty{95}{\percent} confidence, that the true mean rank is no higher than the value used for ranking.

\subsection*{Illustrative example A: two simulation case.}
Suppose only two trials \((\text{num\_simulations}=2)\) are run with a
\(\pm10\,\%\) perturbation of inputs:


\begin{table}[tb]
 \caption{Rank of UCAs after MCS}
    \label{tab:placeholder}
    \centering
    \begin{tabular}{ccc}\toprule
    \textbf{UCA} & $\text{Rank}^{(1)}$ & $\text{Rank}^{(2)}$ \\ \midrule
      \text{UCA-1.1.1} & 1 & 2 \\
\text{UCA-1.2.1} & 2 & 1 \\
\text{UCA-2.1.1} & 3 & 3
    \\ \bottomrule\end{tabular}
   
\end{table}

\begin{table}[tb]
\caption{EJ-Scores in the two simulation example.}
\label{tab:EJ_two_sim}
\centering
\begin{tabular}{lccc}
\toprule
\textbf{UCA} & Average Rank & \(\sigma\) & EJ-Score \\ \midrule
UCA-1.1.1 & 1.50 & 0.50 & 2.00 \\
UCA-1.2.1 & 1.50 & 0.50 & 2.00 \\
UCA-2.1.1 & 3.00 & 0.00 & 3.00 \\ \bottomrule
\end{tabular}
\end{table}
Here in table \ref{tab:EJ_two_sim} the EJ-Score places UCA-1.1.1 and UCA-1.2.1 jointly first.

\subsection*{Illustrative example B: 1000 simulation case.}

 In Table \ref{tab:CI_1000 example}, 1000 trials were run.  
Although the two UCAs again tie on EJ-Score (2.00), the
CI score \(\text{CI}^{\uparrow}_{95}\) breaks the tie
because the larger spread of UCA-1.1.1 is down-weighted by the
\(\sqrt{\text{num\_simulations}}\) term, allowing its lower mean rank to dominate.


\begin{table}
\caption{CI results (1 000 simulations).}
\label{tab:CI_1000 example}
\centering
\begin{tabular}{lrrr}
\toprule
\textbf{UCA} &
EJ-Score &
\(\text{CI}^{\uparrow}_{95}\) &
Final Rank \\ \midrule
UCA-1.1.1  & 2.00 & 1.44 & \textbf{1} \\
UCA-1.2.1  & 2.00 & 1.53 & \textbf{2} \\
UCA-2.1.1  & 3.00 & 3.00 & 3 \\ \bottomrule
\end{tabular}
\end{table}

To prevent embedded uncertainty in the priority list, the CI was combined with the average rank and standard deviation.

\subsection*{Post- MCS review.}


Following the execution of the MCS, a revised ranking of UCAs is obtained. At this stage, a comparative analysis is conducted between the initial expert-based rankings and the MCS-derived rankings. This comparison serves to evaluate the sensitivity of each UCA to changes in the contributing factors, and to derive a final, robust EJ score.
\begin{itemize}
  
\item In some cases, UCAs maintained their initial rank despite minor variations in input values. This consistency indicates that these UCAs are stable, and their priority is robust against uncertainty. 
\item In contrast, some UCAs showed clear changes in their rankings when the input factors were slightly adjusted. These UCAs are considered sensitive, as their priority is strongly influenced by the specific values given to certain factors (the initial Experts’ input). This indicates that the input data for these UCA may require further refinement, and thus a need for another layer (using PMS and CIF) to define their priorities.
\end{itemize}

\begin{figure}[tb]
    \centering
     \includegraphics[width=\linewidth]{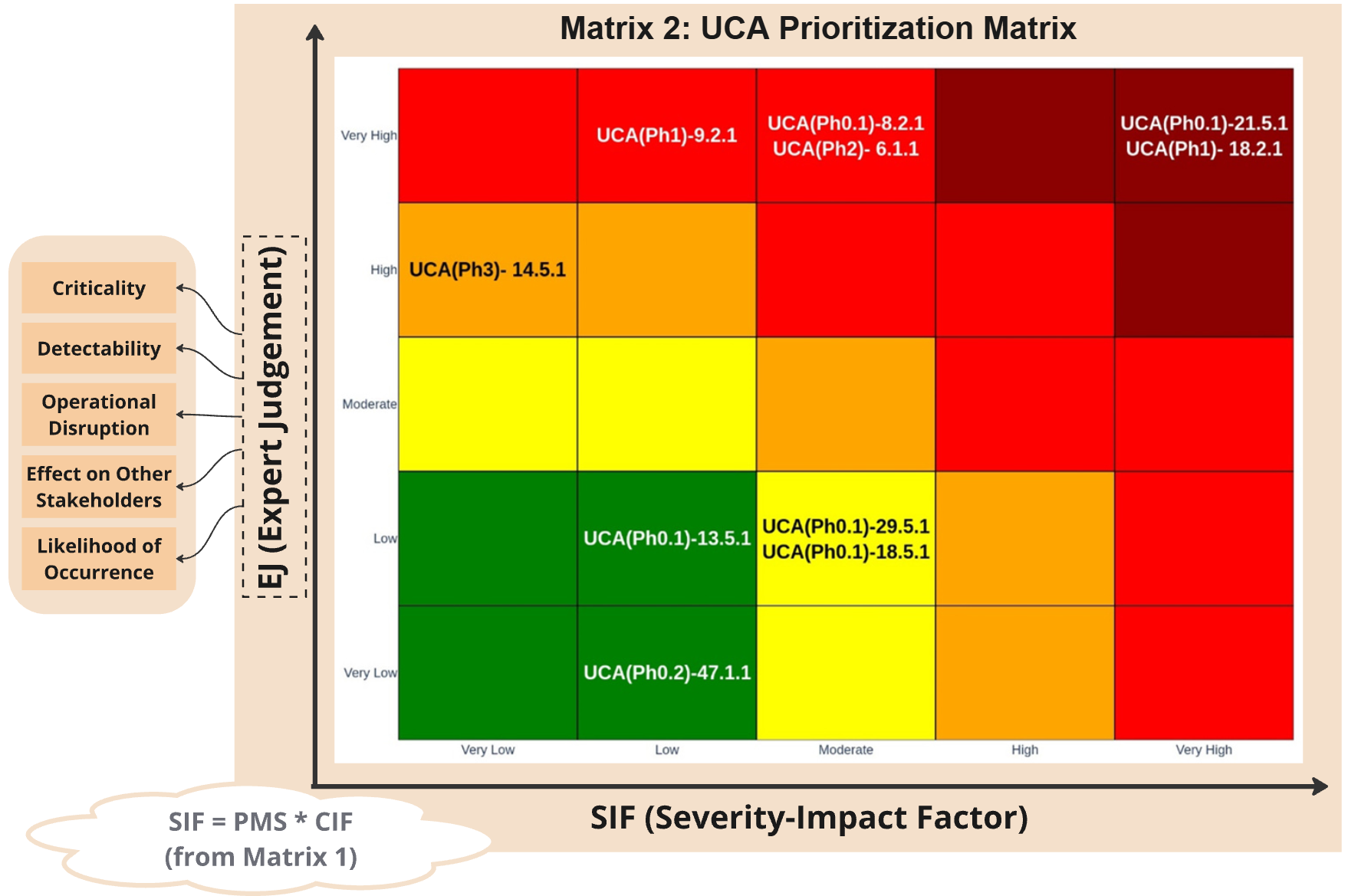}
    \caption{Prioritization Matrix}
    \label{fig:methology mtrix}
\end{figure}

After identifying the required factors to prioritise the UCAs objectively. These factors will assess the UCAs on a matrix, which is called in this paper the \textbf{Prioritisation matrix} \ref{fig:methology mtrix}. 
The next section will depict steps to implement this matrix and populate the results on it.

\subsection{Prioritisation Matrix}
To assess the criticality of the UCAs, it is essential to consider the PMS, CIF and EJ related to each UCA.

\subsubsection{Rules of creating Prioritisation Matrix} 
To create a Prioritisation matrix, there are some rules that need to be followed:
\begin{itemize}
    \item Assign each cell a criticality level based on the provided input data to determine its corresponding level.
    \item Divide the input data into distinct levels, using qualitative descriptions and corresponding scales, respectively.
    \item Scale each input appropriate to UCA in the corresponding cell.
\end{itemize}

 To enable the auto-generation of a matrix, an algorithm was developed in Google Colab to automatically streamline the assignment of UCAs to each cell within the risk matrix.

Normalization of EJ Values with SIF values: As the EJ values are derived from the MCS, it is important to note that, based on the design, the highest importance is allocated to the lowest number. That is why an inverted EJ must be applied to indicate that higher values correlate with greater SIF.
    
\[
\text{EJ\_inverted}_i = \max(\text{EJ}) - \text{EJ}_i
\]

Determining the Range for Each Level: 
The matrix uses five severity levels, represented by colours as follows:
\begin{itemize}
    \item \textbf{Green} (Very Low Priority:P5)
    \item \textbf{Yellow} (Low Priority:P4)
    \item \textbf{Orange} (Minor Priority:P3)
    \item \textbf{Red} (Moderate Priority:P2)
    \item \textbf{Darkred} (High Priority:P1)
\end{itemize}

Given these five categories, each axis (\texttt{SIF} and \texttt{EJ\_inverted}) is divided into five intervals. Let:
\begin{itemize}
    \item \texttt{max\_SIF} be the maximum value of \texttt{SIF}.
    \item \texttt{max\_EJ} be the maximum value of \texttt{EJ\_inverted}.
\end{itemize}

\begin{figure}[tb]
    \centering
    \includegraphics[width=1\linewidth]{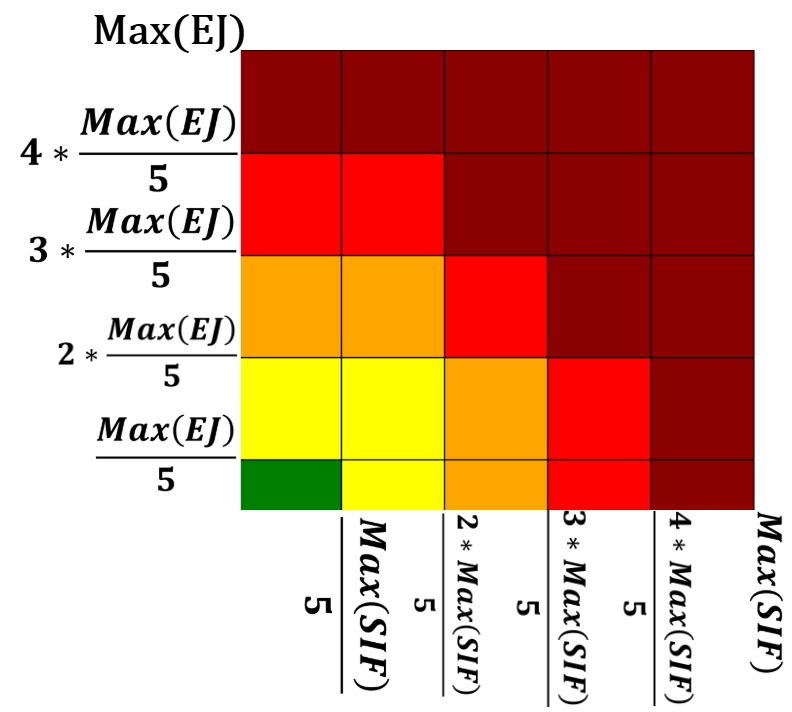}
    \caption{The dynamic scaling of the UCAs.}
    \label{fig:Dyn}
\end{figure}

Each interval will be one-fifth of the maximum value on each axis, yielding the following ranges:
The range for each priority level on both axes is shown in Figure \ref{fig:Dyn}.

Scale SIF and EJ\_normalized to a 5 x 5 grid

\[
\text{sif\_scaled} = \left\lfloor \frac{\text{sif\_value}}{\text{max\_sif}} \times 4 \right\rfloor
\]
\[
\text{ej\_scaled} = \left\lfloor \frac{\text{ej\_normalized\_value}}{\text{max\_ej\_normalized}} \times 4 \right\rfloor
\]

where: \textit{\( \text{sif\_scaled} \)} and \textit{\( \text{ej\_scaled} \)} are the scaled values for \( \text{SIF} \) and \( \text{EJ}_{\text{inverted}} \), respectively, mapped to a 5-level scale (0 to 4).

We set up dynamic scaling by finding the maximum values for \texttt{SIF} and \( \text{EJ}_{\text{inverted}} \). A color gradient to represent varying levels of risk severity is then defined to ensure that the matrix adapts to the dataset’s range without requiring fixed grid dimensions.

\begin{figure*}[tb]
    \centering
     \includegraphics[width=\linewidth]{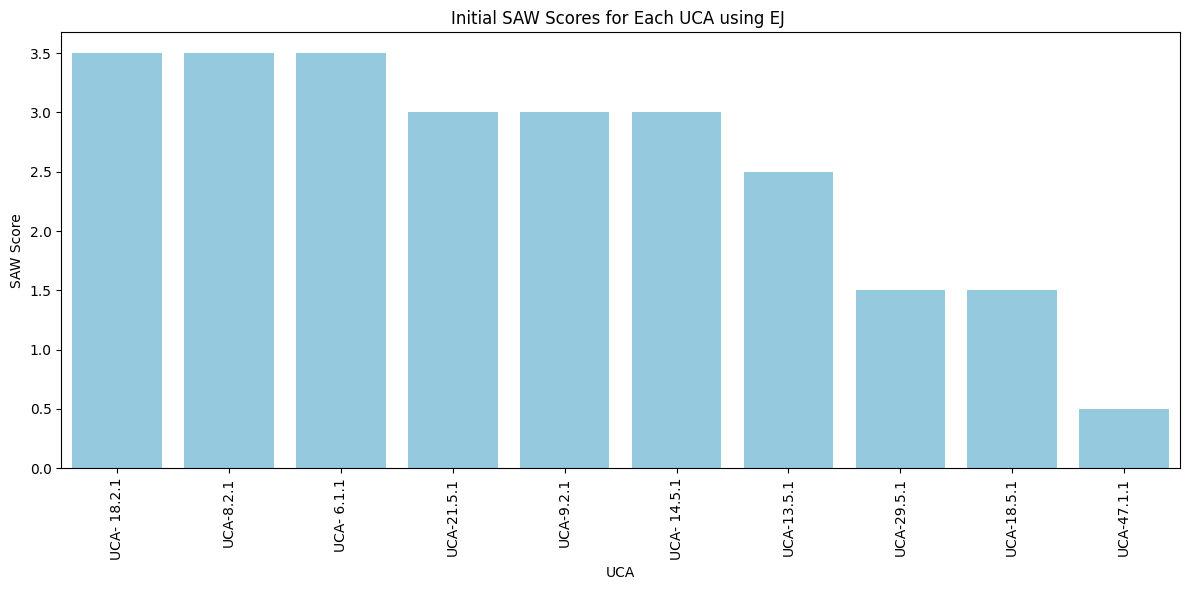}
    \caption{Initial ranking of UCAs based on the EJ}
    \label{fig:SAW based solely on EJ}
\end{figure*}

\begin{figure}[tb]
    \centering
     \includegraphics[width=\linewidth]{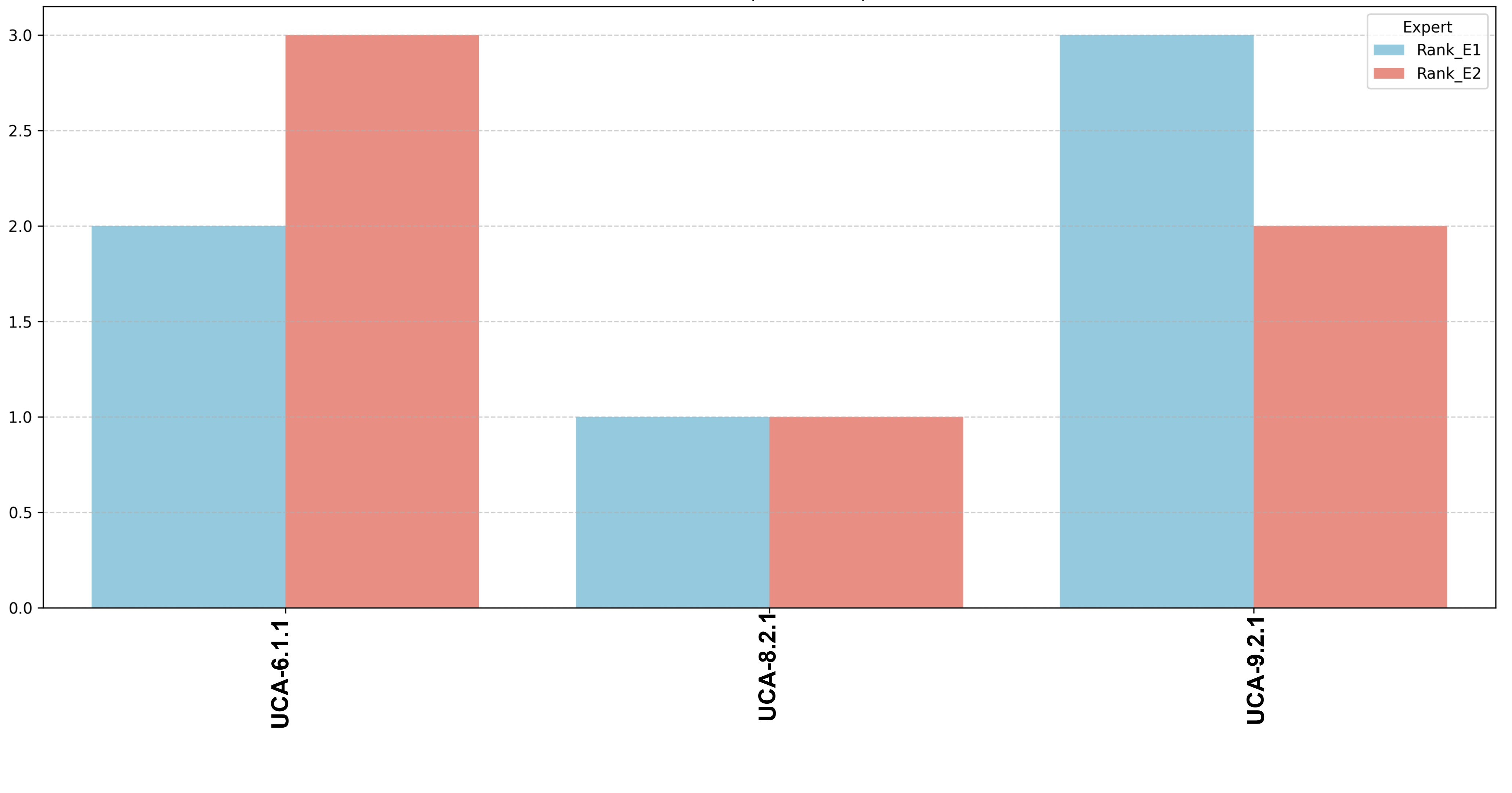}
    \caption{Initial ranking of UCAs based on assessments from two different experts. }
    \label{fig:SAW based on Exp1 and Exp 2}
\end{figure}

\begin{figure*}[tb]
    \centering
     \includegraphics[width=\linewidth]{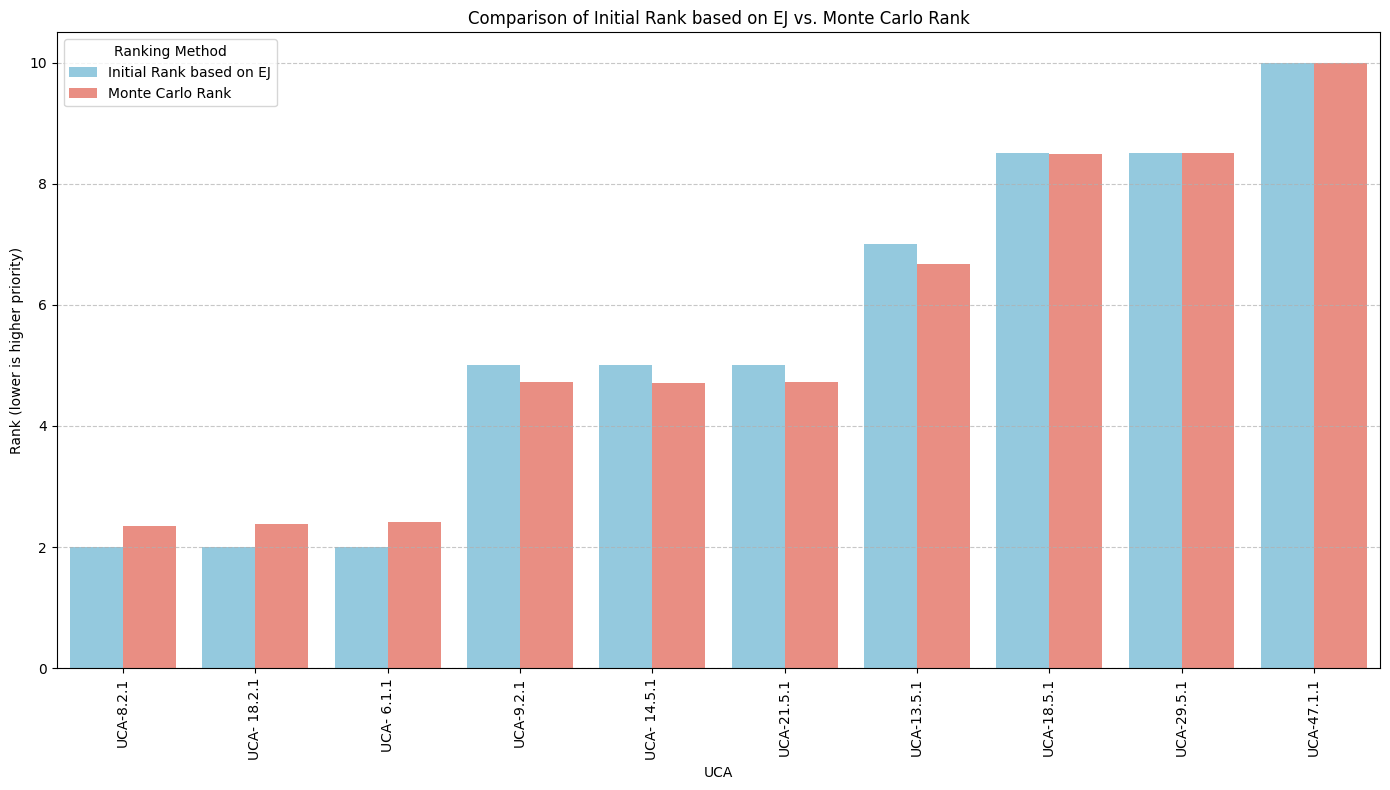}
    \caption{Ranking of UCAs using MCS}
    \label{Ranking of UCAs using MCS}
\end{figure*}

\begin{figure}[tb]
    \centering
     \includegraphics[width=\linewidth]{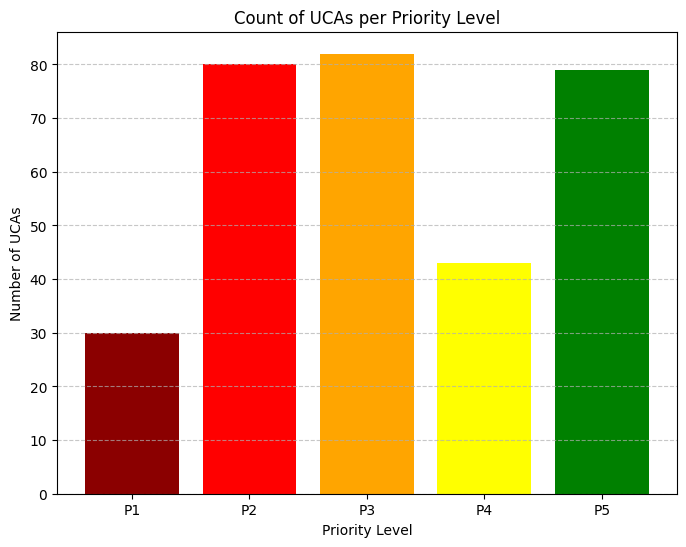}
    \caption{Statistic overview}
    \label{fig:Statistic overview}
\end{figure}

\begin{figure*}[p]
    \centering
     \includegraphics[width=\linewidth]{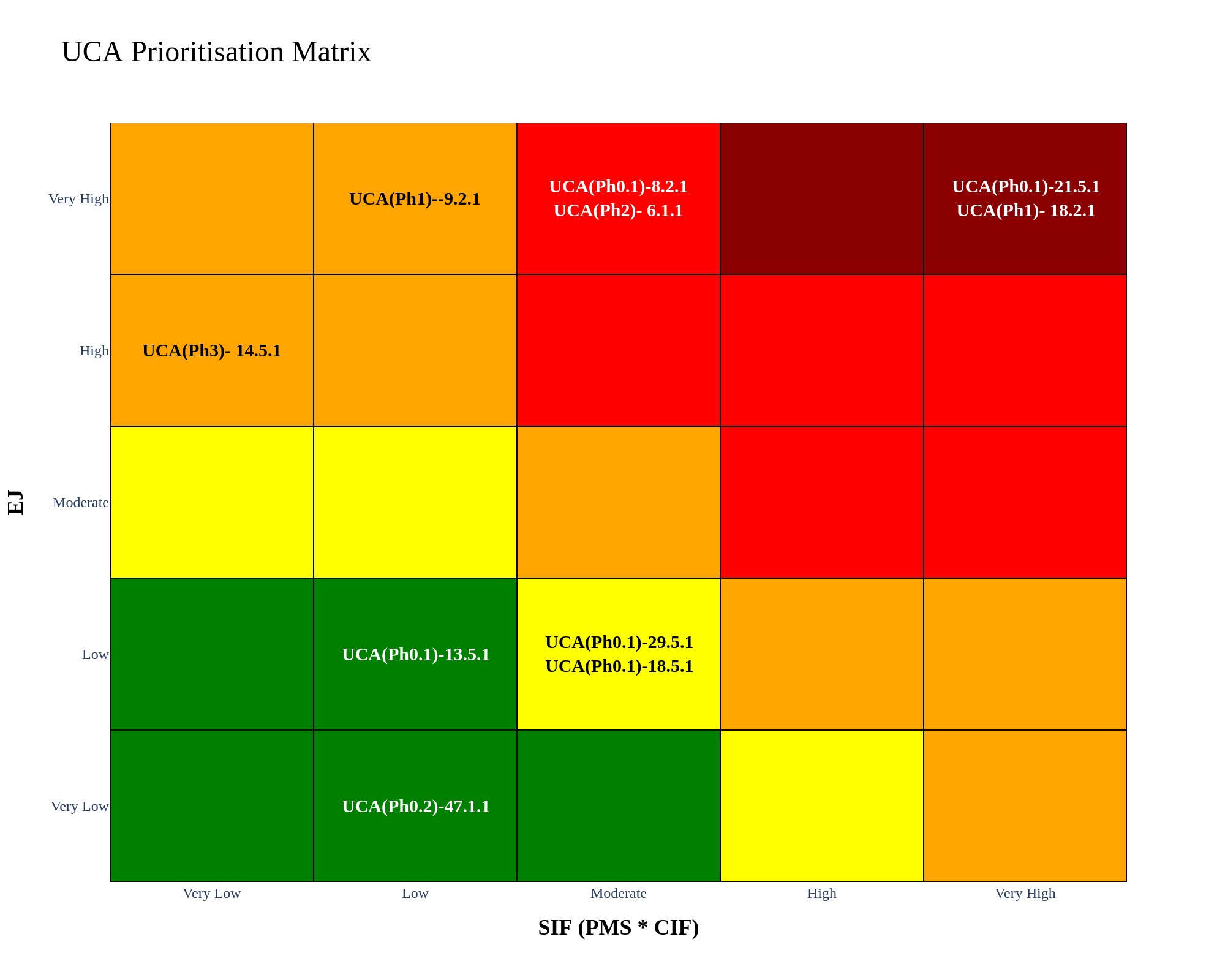}
    \caption{P-matrix for the UCAs used as illustration of the methodology in this paper.}
    \label{P-matrix for 10 UCAs}
\end{figure*}

\begin{table}[tb]
\caption{Table presenting the set of UCAs with their losses}
\label{Tab: L}
\begin{tabularx}{\linewidth}{XXXXXX}
\toprule
\textbf{UCA-ID} & \textbf{L1} & \textbf{L2} & \textbf{L3} & \textbf{L4}& \textbf{L5} \\
\midrule
UCA-21.5.1& L1.1 & L2.1&N/A&N/A&N/A
  \\  \midrule 
UCA- 18.2.1 & L1.1	&L2.1 &	L3.2&	L4.3&	L5.2
 \\  \midrule 
UCA- 8.2.1& L1.1 & L2.1&N/A&N/A&N/A \\
UCA- 6.1.1 & L1.1	&L2.1 &	L3.1&	L4.1&	L5.1 \\  \midrule 
UCA- 9.2.1 & L1.1 & L2.1&N/A&N/A&N/A  \\  \midrule 
UCA- 14.5.1 & L1.1 & N/A &N/A &N/A &N/A \\ \midrule  
UCA- 29.5.1 &N/A &	N/A&	L3.2&	L4.2 &	L5.2
 \\  \midrule 
UCA- 18.5.1& N/A &	N/A&	L3.2&	L4.2 &	L5.2 \\
UCA- 13.5.1& N/A&	N/A	&L3.3&	L4.4&	L5.4
 \\  \midrule 
UCA- 47.1.1&N/A	&N/A&	N/A	&L4.3	&L5.3
\\
\bottomrule
\end{tabularx}
\end{table}

\section{Case Study and Results}
\label{results}

The results of this project were achieved through a collaborative project between WMG and CAA. The Risk Sub-group of the eVTOL Safety Leadership Group (eVSLG) thus decided to apply STPA to understand and establish a new safety analysis procedure for emerging technologies in the aviation industry. The collaboration led to a Civil Aviation Publication (CAP) to support the future integration of eVTOL operation into UK airspace \cite{caa_stpa_evtol}. A total of 318 UCAs were identified from the analysis, followed by the development and application of the UCA prioritisation framework. In this section, 10 UCAs (out of the total 318 UCAs) are prioritised and presented.


\begin{figure*}[tb]
    \centering
     \includegraphics[scale=0.27]{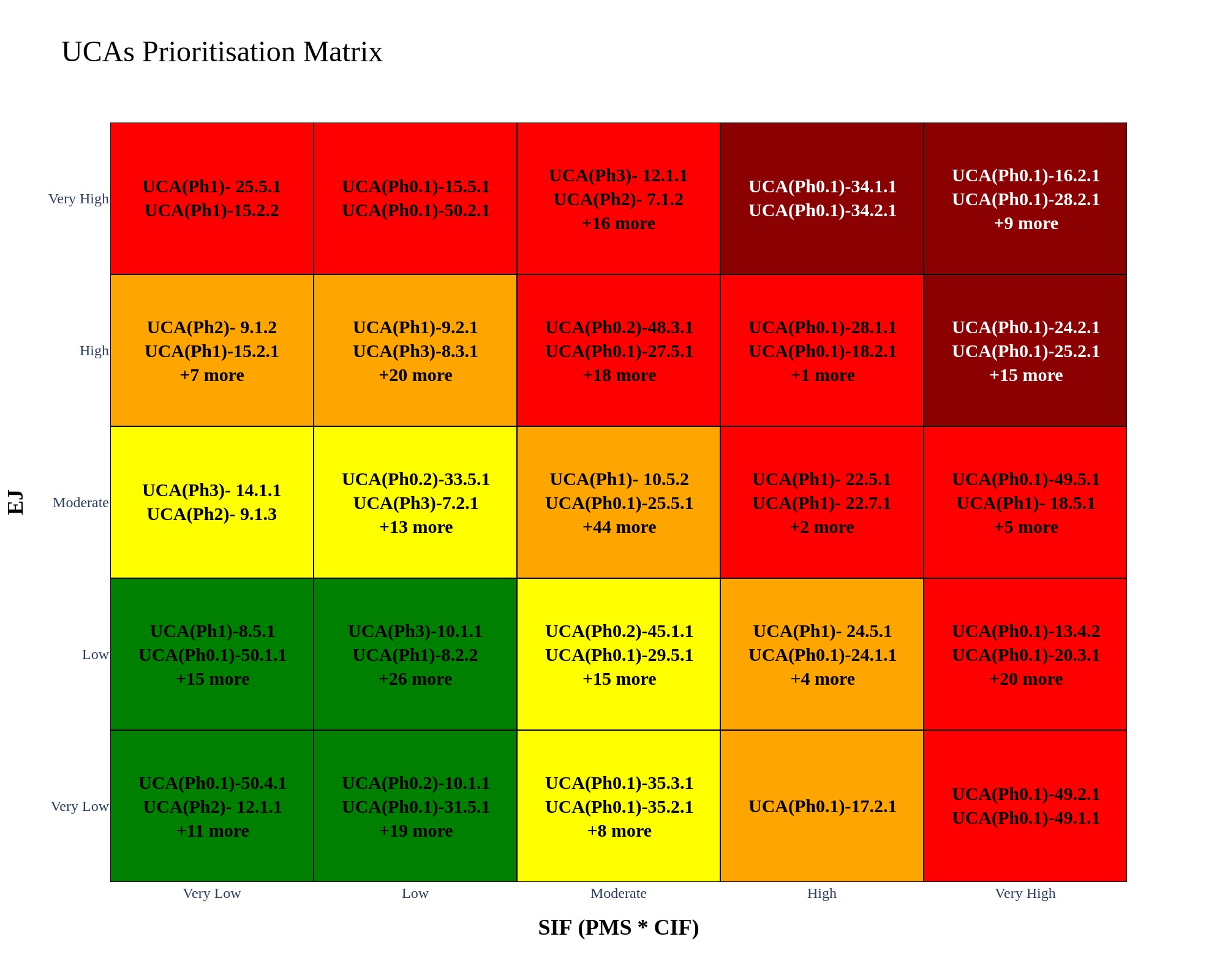}
    \caption{P-Matrix presenting the final ranking of UCAs}
    \label{P-Matrix}
\end{figure*}

Table \ref{Output of STPA Step3:UCAs and their description} presents a set of UCAs that were identified during the STPA process for eVTOL analysis. This table illustrates the methodology discussed in this paper and includes descriptions of each UCA.

\begin{table}[p]
\centering
\caption{Output of STPA Step3:UCAs and their description}
\label{Output of STPA Step3:UCAs and their description}
\begin{tabularx}{\linewidth}{p{1cm}X}
\toprule
\textbf{UCA} & \textbf{Description} \\
\midrule
UCA-21.5.1 & Regulator provides airspace restriction too late when there is already a temporary or permanent modification of airspace usage to ensure safety or security, and the flight is scheduled to fly through that area. \\  \midrule 
UCA- 18.2.1 & NATS (LHR RADAR) provides OnwardClearance incorrectly(incorrect altitude, route,heading and/or speed) when there is a conflict( close proximity to other eVTOLs, helicopters, drones, and traditional aircraft)

  \\  \midrule 
UCA- 8.2.1& Licensed Vertiport (Battersea) provides incorrect/insufficient PTR (Pilot Training Request) when the flight is scheduled, the pilot flying has limited knowledge/experience of the vertiport, and the incorrect PTR is undetected.\\  \midrule 

UCA- 6.1.1 & Licensed Aerodrome does not provide HoldingOutsideATIS when the eVTOL is already at maximum capacity and cannot accommodate additional eVTOLs in holding patterns.
Note : This creates a risk of overcrowding in the holding area, which increases the likelihood of mid-air collisions due to insufficient space between eVTOLs. \\  \midrule 
UCA- 9.2.1 & The eVTOL Operator provides incorrect FlightChecks (mass and balance calulations) and the flight takes off\\  \midrule 

UCA- 14.5.1 & Commander provides Landing Request too late after  the eVTOL is descending towards a busy aerodrome under low-visibility conditions. 

 \\  \midrule 
UCA- 29.5.1 & NATS (LHR RADAR) provides LOA (Letter of Agreement) too late when the flight is already being planned.
Note: this would delay the progress of flight operation planning, leading to business-critical losses."
 \\  \midrule 
UCA- 18.5.1& Regulator provides COA (Certificate of Authorization) too late when the aircraft has met the safety standards, it has been registered, and it is in schedule for the flight operation, and there are no alternative aircraft available.
 \\  \midrule 
UCA- 13.5.1& Regulator provides Approval too late (by x weeks) when the flight is already scheduled.
\\  \midrule 
UCA- 47.1.1& Licensed Vertiport (Battersea) does not provide safety instructions when passengers boarding have very limited knowledge/awareness of the safety process.
 \\
\bottomrule
\end{tabularx}
\end{table}

To derive the PMS, as explained in the methodology section, we follow STPA step 3, which involves identifying UCAs. Each UCA must be linked to at least one potential loss. As part of this project, sub-losses were identified and allocated to each UCA, as shown in \cref{Tab: L}.


\begin{table*}[tb]
\centering
\caption{Experts evaluation of the UCAs}
\label{Experts evaluation to the UCAs}
\begin{tabularx}{\linewidth}{lXllXX}
\toprule
\textbf{UCA-ID} & \textbf{Operational Disruption} & \textbf{Criticality } & \textbf{Detectability} & \textbf{Effect on Other Stakeholders}&\textbf{Likelihood of occurrence} \\
\midrule
UCA-21.5.1& High Impact& High Risk& High Detectability  &Significant Impact& 1
  \\  \midrule 
UCA- 18.2.1 & Medium Impact & High Risk &Low Detectability &Significant Impact& 1 \\  \midrule 
UCA- 8.2.1& High Impact & High Risk & Moderate Detectability &Significant Impact& 1 \\  \midrule 
UCA- 6.1.1 & High Impact & High Risk & Moderate Detectability & Significant Impact&1 \\  \midrule 
UCA- 9.2.1 & Medium Impact & High Risk & Low Detectability& Moderate Impact&1 \\  \midrule 
UCA- 14.5.1 & High Impact & High Risk& High Detectability& Significant Impact&1 \\  \midrule 
UCA- 29.5.1 & Medium Impact &Moderate Risk & High Detectability&Moderate Impact&1 \\  \midrule 
UCA- 18.5.1& High Impact  & Moderate Risk & High Detectability&Minimal Impact& 1 \\  \midrule 
UCA- 13.5.1& High Impact  & Moderate Risk & High Detectability& Significant Impact&0 \\ \midrule 
UCA- 47.1.1& Low Impact & Low Risk & Moderate Detectability &Minimal Impact& 1 \\
\bottomrule
\end{tabularx}
\end{table*}

\begin{table}[tb]
\centering
\caption{Results table with SIF and EJ}
\label{Tab:Results table with SIF and EJ}
\begin{tabularx}{\linewidth}{lcccX}
\toprule
\textbf{UCA-ID} & \textbf{PMS} & \textbf{CIF} & \textbf{EJ} & \textbf{Likelihood of occurrence} \\
\midrule
UCA-21.5.1& 20 & 6 & 59.4072555&1
  \\  \midrule 
UCA- 18.2.1 & 20 & 5 & 29.85918475 & 1 \\  \midrule 
UCA- 8.2.1& 20 & 4 & 29.77339235 & 1 \\  \midrule 
UCA- 6.1.1 & 20 & 3 & 29.87185488 & 1 \\  \midrule 
UCA- 9.2.1 & 20 & 2 & 58.99621273& 1 \\  \midrule 
UCA- 14.5.1 & 20 & 1& 59.45807616& 1 \\ \midrule 
UCA- 29.5.1 & 12 & 5 & 208.2534994& 1 \\  \midrule 
UCA- 18.5.1& 12 & 6 & 208.6651534& 1 \\  \midrule 
UCA- 13.5.1& 4 & 6 & 208.8436053& 0 \\  \midrule 
UCA- 47.1.1& 7 & 4 & 266.8445137& 1 \\
\bottomrule
\end{tabularx}
\end{table}

Table \ref{Experts evaluation to the UCAs} presents the EJ rankings, which were collected from the SMEs from different organisations during a series of workshops throughout the project timeline.

Table \ref{Tab:Results table with SIF and EJ} presents the three factors required to implement the methodology: PMS, CIF, and EJ, the last of which was subjected to a MCS to reduce the uncertainty inherent in the experts' scores.


The initial ranking of UCAs, based on the experts' inputs, is presented in Figure \ref{fig:SAW based solely on EJ}, following the SAW calculation process illustrated in Figure \ref{fig:EJ score Calculation Process} and indicated by the red arrows.


The results of applying a MCS are presented in Figure \ref{Ranking of UCAs using MCS}, showing a comparison between the experts' judgments accounting for uncertainty and the MCS outputs, which aim to reduce that uncertainty. To ensure an accurate comparison, the SAW values were inverted to align with the MCS output, as MCS assigns higher priority to lower numerical values. 

Once all three metrics: PMS, CIF, and EJ have been collected and assigned, a Prioritisation Matrix is used, as shown in Figure \ref{P-Matrix}, to present the final ranking across five levels (Use dark red to indicate high priority and green for the lowest priority.) reflecting the priority of each UCA.

 \section{Discussion}
 
In this paper, a UCA prioritization framework is introduced to more objectively prioritize the UCAs identified from STPA. The prioritized sets of UCAs are captured in the UCA prioritization matrix (Figure \ref{P-matrix for 10 UCAs}) for better communication of the results. The prioritization matrix is color-coded based on the criticality level. In the case study, 10 UCAs were evaluated, and 4 were identified as high-priority based on the factors incorporated during the analysis: the PMS, CIF, and EJ.
In this section, we interpret the results presented in Section \ref{results}. First, we collected expert judgments for the UCAs (see \ref{Experts evaluation to the UCAs}). To prepare our data for MCS evaluation, these judgments were converted into scores, as explained in Section \ref{EJ}. Next, we applied the SAW method according to Equation \ref{SAW} to obtain an initial ranking based on expert judgments. However, because expert judgments can be subjective, we conducted an MCS to reduce this subjectivity, following the methodology described in Section \ref{EJ}. As shown in Figure \ref{fig:SAW based solely on EJ}, the first three UCAs have identical scores, which does not provide a basis for prioritizing mitigation actions and leaves room to question the certainty and objectivity of the ranking.

In Figure~\ref{fig:SAW based on Exp1 and Exp 2}, the uncertainty and subjectivity inherent in expert rankings are illustrated. The figure highlights differences in the ranking of the same UCAs by two independent experts. Notably, a rank reversal between UCA-6.1.1 and UCA-9.2.1 is observed. This discrepancy underscores the uncertainty in outcomes, this is mainly due to variations in each expert’s mental models, judgment criteria, and professional experiences.

Given this observed uncertainty , there is a clear need for an approach capable of mitigating expert subjectivity and reducing uncertainty in safety assessments. MCS provides this capability. By performing extensive simulations across random variations, MCS delivers outputs that are less subjective to personal bias. and more robust to uncertainty. The resulting of this simulation is depicted in Section \ref{Ranking of UCAs using MCS}. The graph in figure \ref{Ranking of UCAs using MCS} shows a slight gap between the initial ranking and the MCS, indicating moderately higher sensitivity to parameter variations.

Specifically, UCAs-29.5.1 (i.e., NATS (LHR RADAR) provides LOA (Letter of Agreement) too late when the flight is already being planned), UCA-18.5.1 (i.e., Regulator provides COA (Certificate of Authorization) too late when the aircraft has met the safety standards, it has been registered, and it is in schedule for the flight operation, and there are no alternative aircraft available), and UCA-47.1.1 (i.e., Licensed Vertiport (Battersea) does not provide safety instructions when passengers boarding have very limited knowledge/awareness of the safety process), are consistently ranked as low critical in both the SAW and MCS evaluations. This consistency suggests that despite the small parameter variations, they initially appear low risk. However, to ensure their non-critical status, an additional layer of analysis, integrating MCS and other factors (PMS and CIF) is applied to confirm the final risk classification.

Conversely, the wide error gap between the SAW, as presented in Figure \ref{fig:SAW based solely on EJ} and MCS in Figure \ref{Ranking of UCAs using MCS} approaches for the other UCAs indicates that more refined data are needed to spot the criticality effectively, and expert inputs alone are insufficient for accurate ranking.

When safety is paramount, objectivity is essential in ranking these UCAs. While reducing the number of UCAs can help manage the analysis outputs, it is crucial not to overlook any UCA that might lead to catastrophic losses. Therefore, we added an additional layer to the prioritisation process by incorporating severity factors, the PMS and CIF (impact factor). As shown in the matrix (Figure \ref{P-Matrix}), the final ranking of the UCAs is presented alongside their associated factors: the SIF and the EJ score.

Recognizing that safety is not an exact science, we opted for a matrix presentation that depicts criticality across five levels, from very low to very high instead of using unique numbers. The two axes of the P-Matrix (SIF and EJ) represent complementary views. The colour-coded presentation helps focus on areas of criticality and identify UCAs that require urgent intervention. For example, UCAs 18.2.1 (i.e., NATS (LHR RADAR) provides OnwardClearance incorrectly(incorrect altitude, route, heading and/or speed) when there is a conflict( close proximity to other eVTOLs, helicopters, drones, and traditional aircraft) and 21.5.1 (i.e., Regulator provides airspace restriction too late when there is already a temporary or permanent modification of airspace usage to ensure safety or security, and the flight is scheduled to fly through that area) are located in the dark-red area, indicating they are very critical and require immediate mitigation measures. These two UCAs are linked to catastrophic losses (e.g., loss of life), and, as defined by stakeholders, such losses are unacceptable. Based on the CIF, these UCAs have a very high impact factor, and failing to mitigate them could affect other controllers and lead to severe losses. Stakeholder input also confirms that these are highly relevant and likely to occur, justifying their placement in the dark-red area and their high-priority status. Since the CIF is structurally assigned at the controller level, it might not accurately represent the true influence or downstream impact of specific control actions. Future research could improve CIF by assessing the impact at the level of individual control actions, instead of applying a uniform CIF to all Control Actions from a given Controller. Alternatively, analyzing the number of affected control actions issued by a Controller could yield more detailed insights, although this approach may require greater computational resources.

In contrast, the UCAs in the green area of Figure \ref{P-Matrix} are estimated to be non-critical. For instance, UCAs 13.5.1 (i.e., Regulator provides Approval too late (by x weeks) when the flight is already scheduled) and 47.1.1 (i.e., Licensed Vertiport (Battersea) does not provide safety instructions when passengers boarding have very limited knowledge/awareness of the safety process) are associated with partial tactical mission losses and major loss in customer satisfaction. Experts have also ranked these UCAs as very unlikely to occur, which is reflected in their EJ scores derived from the MCS, thereby justifying their placement in the green area.

Implementing this approach has yielded promising results in effectively ranking the UCAs. By integrating the tri-metric parameters (PMS, CIF, and EJ scores), the resulting prioritisation matrix is both analytically rigorous and visually self-explanatory. This integration combines two complementary perspectives: the STPA analyst’s evaluation, which assesses each UCA’s potential link to losses and its position within the control structure hierarchy, and expert judgment, which captures specialists' perceptions of risk based on their knowledge and experience. Consequently, this structured approach enables experts to assign meaningful scores to UCAs by evaluating their context and likelihood of occurrence, while the STPA analysis quantifies the potential impact of each UCA on overall system safety.

Critical UCAs identified within the dark-red area of the prioritisation matrix (see Figure~\ref{fig:Statistic overview}) are clearly associated with catastrophic losses deemed unacceptable by stakeholders. In contrast, UCAs positioned in the yellow or green areas typically correlate with non-catastrophic business losses. Moreover, integrating STPA insights with expert judgment and validating these assessments through Monte Carlo Simulation (MCS) has significantly reduced uncertainty, generating robust data that effectively addresses research questions RQ1 and RQ2. The resulting prioritisation matrix mirrors the logic of a conventional risk matrix, clearly depicting risk criticality by severity (represented by the SIF factor) and likelihood (captured by the expert judgment score).


In the absence of a standard scientific framework for prioritization of STPA results, the project stakeholders found this approach valuable for facilitating communication and early decision making, especially considering the tight timeline for the introduction of eVTOLs \cite{caa_stpa_evtol}.

Dynamic scaling allows for data scaling in the prioritisation matrix based on the priority level in the dataset. We opted for this option to enforce and ensure that each UCA is assigned to the appropriate priority level that correlates with the three factors: PMS, CIF, and EJ. Also, address the research question RQ3. The input into this methodology consisted of 317 UCAs. The scaling of this data (refer to figure \ref{P-Matrix}) is not the same as when analysing the 10 UCAs used to illustrate this methodology in this paper\ref{P-matrix for 10 UCAs}.

The MCS is a widely implemented simulation across various domains and has proven effective in reducing uncertainty in expert judgments. When its output is combined with the PMS and CIF factors, the results become robust for ranking critical UCAs. and helped manage hundreds of UCAs generated by the STPA paradigm without compromising the study’s primary objective: preventing any unacceptable losses as defined by stakeholders.
This approach has been implemented in a real-world case study conducted in collaboration with the UK Civil Aviation Authority (CAA), where the Regulator intends to evaluate how the study’s findings, outlined in the report \cite{caa_stpa_evtol}, can be integrated into their regulatory work programme.

\section{Conclusion}
This paper presents a novel methodology that addresses a limitation of STPA, allowing for the effective management of a large number of inputs without overlooking potential UCAs that might lead to catastrophic losses. Our proposed approach does not aim to reduce the number of UCAs but primarily focuses on identifying the most relevant ones that require urgent intervention and mitigation to prevent catastrophic losses.

This methodology was implemented in a real-world aviation project, and the results are promising. They successfully highlighted the most critical UCAs across different departments Out of 318 UCAs, 110 have been ranked as high priority.

Integrating valuable insights from domain experts has contributed to defining specific contexts in which these UCAs might occur. Additionally, applying MCS in the process has reduced subjectivity in these assignments. Incorporating the new EJ score alongside PMS and CIF has resulted in a more robust UCAs ranking.

The methodology is not claimed to be perfect, and further application in real-world case studies is necessary to evaluate its effectiveness and identify potential improvements.

As part of our future work, we aim to extend this approach across different case studies and domains to enhance its applicability and robustness. Also improve the interpretability and construction of the CIF Metric. A supporting tool will also be implemented to facilitate the execution of the methodology. Building on the same motivations to achieve this work, our next step is to explore requirements prioritisation and management to enhance the fourth step of STPA.

\section{Acknowledgements}
The work presented in this paper was funded by the UK Civil Aviation Authority (GFA 3549). The authors would also like to thank the WMG center of HVM Catapult and WMG, University of Warwick, UK, for providing the necessary infrastructure for conducting this study. WMG hosts one of the seven centers that together comprise the High-Value Manufacturing Catapult in the UK.

\bibliographystyle{elsarticle-num} 
\bibliography{root}

\end{document}